\documentclass[final,review,3p,times]{elsarticle}

\usepackage[version=3]{mhchem}
\usepackage{subfig}
\usepackage{graphicx}
\usepackage[table]{xcolor}
\usepackage{array}
\usepackage{ulem}
\usepackage{float}
\usepackage{amsthm}  
\usepackage{amsmath} 
\usepackage{amssymb} 
\usepackage{mathabx} 
\usepackage{mathtools} 
\usepackage{multirow}
\usepackage{rotating}
\usepackage{textcomp}
\usepackage{float}
\usepackage{color}
\usepackage[colorinlistoftodos]{todonotes}
\usepackage{placeins}
\usepackage{ecrc}
\usepackage{longtable}
\usepackage{booktabs}				
\usepackage{tabularx}
\usepackage{placeins}
\usepackage{lscape}
\usepackage{todonotes}
\usepackage{eurosym}
\usepackage{url}
\usepackage{siunitx}  

\volume{00}
\firstpage{1}
\journalname{Journal of Membrane Science}
\runauth{D. Rall et al.}
\jid{J. Memb. Sci.}
\jnltitlelogo{J. Memb. Sci.}

\begin{document}
\begin{frontmatter}


\title{Multi-scale membrane process optimization with high-fidelity ion transport models through machine learning}


\author[AVT.CVT,DWI]{Deniz Rall\corref{authors}}
\author[AVT.SVT]{Artur M. Schweidtmann\corref{authors}}
\author[AVT.CVT]{Maximilian Kruse}
\author[AVT.CVT]{Elizaveta Evdochenko}
\author[AVT.SVT]{Alexander Mitsos}
\author[AVT.CVT,DWI]{Matthias Wessling\corref{mycorrespondingauthor}}

\cortext[authors]{These authors contributed equally to this work}
\cortext[mycorrespondingauthor]{Corresponding author: manuscripts.cvt@avt.rwth-aachen.de}

\address[AVT.CVT]{RWTH Aachen University, AVT.CVT - Aachener Verfahrenstechnik, Chemical Process Engineering, \\ Forckenbeckstrasse 51, 52074 Aachen, Germany}
\address[DWI]{DWI - Leibniz Institute for Interactive Materials, \\ Forckenbeckstrasse 50, 52074 Aachen, Germany}
\address[AVT.SVT]{RWTH Aachen University, AVT.SVT - Aachener Verfahrenstechnik, Process Systems Engineering, \\ Forckenbeckstrasse 51, 52074 Aachen, Germany}

\begin{abstract}
Innovative membrane technologies optimally integrated into large separation process plants are essential for economical water treatment and disposal.
However, the mass transport through membranes is commonly described by nonlinear differential-algebraic mechanistic models at the nano-scale, while the process and its economics range up to large-scale. Thus, the optimal design of membranes in process plants requires decision making across multiple scales, which is not tractable using standard tools. 
In this work, we embed artificial neural networks~(ANNs) as surrogate models in the deterministic global optimization to bridge the gap of scales. This methodology allows for deterministic global optimization of membrane processes with accurate transport models -- avoiding the utilization of inaccurate approximations through heuristics or short-cut models.
The ANNs are trained based on data generated by a one-dimensional extended Nernst-Planck ion transport model and extended to a more accurate two-dimensional distribution of the membrane module, that captures the filtration-related decreasing retention of salt. We simultaneously design the membrane and plant layout yielding optimal membrane module synthesis properties along with the optimal plant design for multiple objectives, feed concentrations, filtration stages, and salt mixtures.
The developed process models and the optimization solver are available open-source, enabling computational resource-efficient multi-scale optimization in membrane science. 
\end{abstract}




\end{frontmatter}

\sloppy{}
%
%
%
%
%
%
%
%
%
%
\newpage
\section{Introduction}

Global population growth and increasing urbanization require novel economically viable concepts for drinking water treatment and water disposal~\cite{larsen2016emerging}. This challenge includes smart operations, such as data-driven urban water management~\cite{eggimann2017potential}, but also new treatment concepts and technologies~\cite{al2019can, bagheri2019advanced}. Synthetic membranes are an essential technological basis for this separation processes~\cite{nunes2019thinking, ghaffour2013technical}. Innovative membrane technologies help to develop new approaches for the reuse of sustainable raw material sources from urban and industrial waste water~\cite{abels2013membrane, niewersch2014nanofiltration} as well as producing smart process water tailored to its application~\cite{nair2018membrane}. These ambitious aspirations require a versatile modeling environment and rigorous optimization methodologies for membrane systems made for case-specific customized processes. 
 
Nowadays, ion selectivity becomes increasingly essential for sustainable drinking water treatment and electrochemical processes~\cite{werber2016materials, luo2018selectivity}. A transition followed from the complete removal of all minerals from drinking water through reverse osmosis membranes towards more selective nanofiltration membranes. The selectivity of ionic components enables the recovery of sustainable raw material sources from urban and industrial waste water~\cite{shannon2008science}, such as phosphorus~\cite{remmen2019phosphorus}. 
The development of ion separation membranes is driven by the advent of the layer-by-layer (LbL) nanofiltration technology~\cite{liu2018porous, harris2000layered, malaisamy2005high}. The layer-by-layer (LbL) nanofiltration technology is characterized by oppositely charged polymer layers applied to a conventional highly porous ultrafiltration membrane. The layer-wise manufacturing technique enables precise tuning of selectivity and permeability~\cite{cheng2018selective, ilyas2017preparation, menne2016regenerable} and a scale-up from the laboratory to industrial scale~\cite{menne2016precise}.
However, tailoring of the selectivity of charged and uncharged species \cite{rall2019rational, labban2018relating, dirir2014theoretical} is the predominant challenge. Moreover, exploiting all synergies in membrane development through the simultaneous design of LbL membranes, the process structure, and its operation is an open research question~\cite{rall2020simultaneous}.

\begin{figure}[H]
  \centering
  \includegraphics[width=0.75\textwidth]{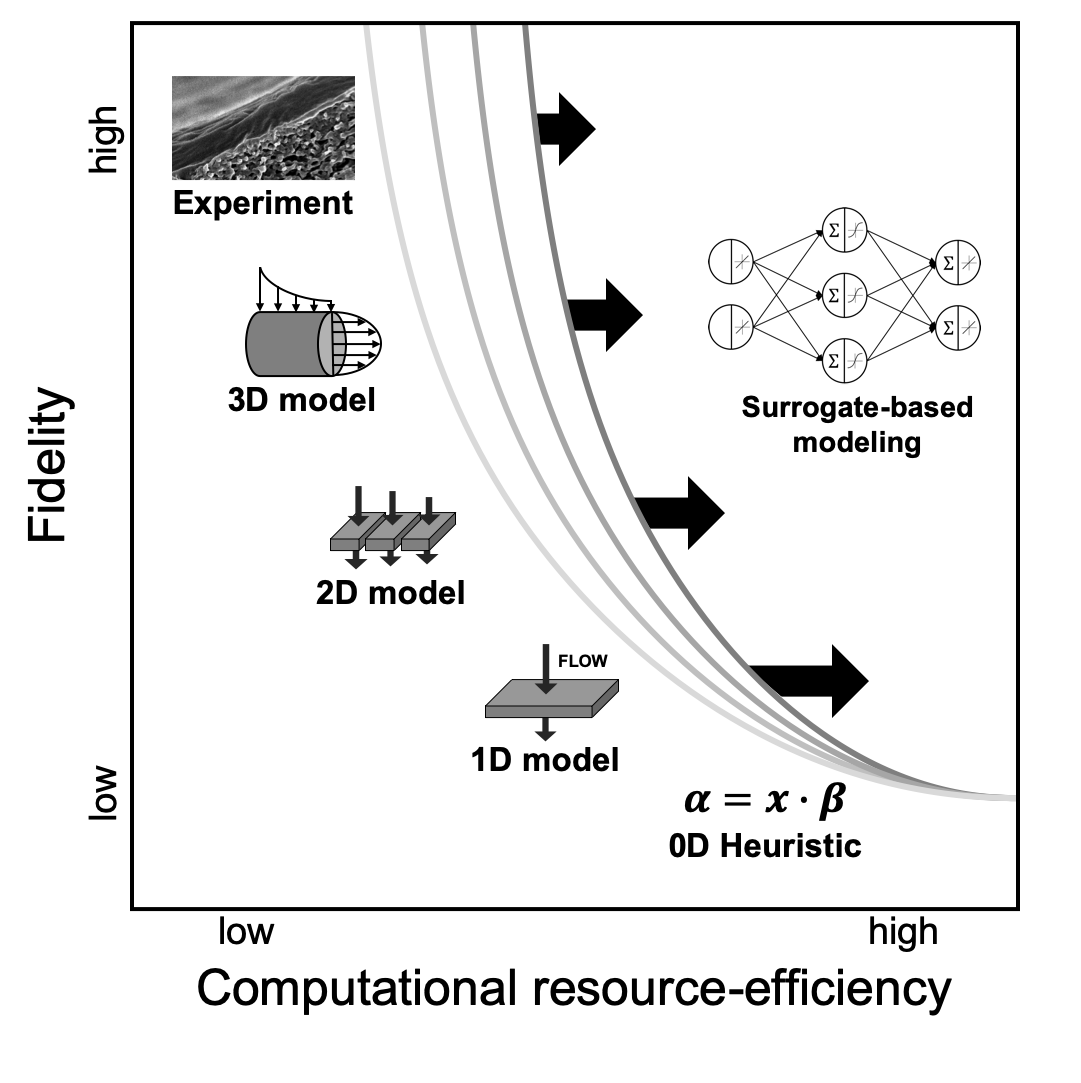}
        \vspace{-1cm}
        \caption{ANN-based surrogate models push the boundary of computational recource-efficiency: Fidelity and computational recource-efficiency trade-off of different descriptive approaches. The descriptive approaches include experiments, simulation studies, and heuristics. Adapted from~\cite{jin2011surrogate}}
        \label{fig:Overview}
\end{figure}

Detailed mechanistic models describe the mass transport and separation by membranes at the nano-scale. There exists a variety of different membrane models for ion separation~\cite{lonsdale1965transport, schlogl1966membrane, yaroshchuk2013solution, femmer2016mechanistic, bowen2002modelling}. These mechanistic models range from highly accurate 3D models to low fidelity heuristics, as shown in Figure~\ref{fig:Overview}. Most notably, a trade-off exists between the fidelity of the approach and its 3D computational resource-efficiency~\cite{jin2011surrogate}. 
Accurate mechanistic models describe ion transport mechanisms in multiple dimensions, leading to partial differential equations (PDE) that require special solvers for the solution, i.e., computational fluid dynamics~(CFD) for 3D. Lumping dimensions reduces the dimensionality to 2D and 1D models that can be evaluated at lower computational effort at the cost of fidelity. Finally, heuristics describe lumped systems (0D) and can be assessed at little computational effort, but their predictive capability is insufficient to describe the complex ion transport. Moreover, only experiments obtain high fidelity data directly.
This results in the dilemma that increased accuracy leads to enormous computational effort and cannot be used in the context of process optimization by using standard tools. 

Economically viable and sustainable process design requires decision making at large process scales (potentially also considering environmental effects~\cite{lapkin2010chemical}). At the same time, physical models of the complex ion transport are described at small-scales. This multi-scale problem has strong implications for optimal decision making:
On the one hand, superstructure process optimization with 3D, 2D, and 1D models embedded lead to (mixed-integer) optimization problems with PDE constraints. These mixed-integer dynamic optimization~(MIDO) problems~\cite{chachuat2006global,singer2006global,wesselhoeft2018algorithms} are very difficult-to-solve and their deterministic global optimization is currently limited to small-scale problems~\cite{sager2015efficient}. Thus, process optimization using this approach is currently not possible.
On the other hand, heuristics or short-cut models can be embedded computational resource-efficiently in complex optimization superstructures~\cite{ohs2016optimization, zarca2019optimization, lee2018automated, bocking2019can, mores2018membrane, alsayegh2017systematic, ghobeity2014optimal}. However, these simplified heuristics or short-cut models do not sufficiently describe the influence of membrane parameters, such as the synthesis protocols or process environment, on the membrane performance. Simultaneous development of the membrane synthesis, along with the process using heuristics or short-cut models, is not possible. 

Overcoming the trade-off between fidelity and computational resource-efficiency is necessary for deterministic global membrane process optimizations with the simultaneous development of the membrane synthesis.
One way to overcome this issue is to replace the expensive simulation by a surrogate model~\cite{white2019multiscale}. Here, the accurate model is evaluated offline, creating a training data set that is subsequently learned by a supervised machine learning algorithm. This results in a data-driven surrogate model that approximates the accurate model. Then, the data-driven surrogate model can be combined with further data-driven and mechanistic models yielding a hybrid mechanistic / data-driven model that can be optimized~\cite{von2014hybrid}. The use of machine learning models enables bridging scales, e.g., in material design~\cite{zhou2019big, tsay2019110th, prakash2018chances, sanchez2018inverse, henao2011surrogate, unger2009neural} or process systems engineering~\cite{lee2018machine,venkatasubramanian2019promise}.
In addition, data-driven models have been used in various disciplines for process optimization~(e.g.,~\cite{mistry2018mixed,boukouvala2016data,cozad2014learning,fahmi2012process,del2019review,schafer2019reduced}).
Further, we have found that hybrid mechanistic / data-driven models with artificial neural networks~(ANNs) embedded can be optimized efficiently by a reduced-space deterministic global optimization method~\cite{schweidtmann2019deterministic}. In the past, this new optimization approach has already been used successfully for the optimization of energy processes~\cite{schweidtmann2019singlespecies,huster2019WorkingFluidSelection,schweidtmann2019flash,huster2019impact,schafer2019wavelet}.

Many publications in the field of membrane science address the implementation of neural networks in membrane development~\cite{madaeni2010modeling, al2007rejection, wessling1994modelling} or prediction of membrane operation~\cite{roehl2018modeling, salehi2016modeling, soleimani2013experimental}. Recently, we extended the use of ANN surrogate models in membrane science to describe and optimize LbL-based membrane systems by identifying superior membrane synthesis protocols based on the delicate trade-off between ion retention characteristics and permeability~\cite{rall2019rational}. 
Next, a coupling of the ANNs into a hybrid mechanistic / data-driven model enables an optimization strategy to simultaneously design the performance of LbL nanofiltration membrane modules and the separation process. This yields membrane synthesis protocols and membrane processes that are optimally tailored to the desired separation task~\cite{rall2020simultaneous}. The results suggest that simultaneous membrane synthesis and process optimization design achieve immediately favorable results with lower impurities at comparable costs. 
But in the previous work, this optimization has been limited to lab data. In particular, only a single-stage membrane process has been considered because the available experimental data is only valid for a fixed feed concentration and single salt solutions.
Thus, an extension to accommodate for higher feed concentrations, multiple-staged processes, and salt mixtures is highly desired.

In this work, we propose a surrogate-based approach to bridge the gap between mechanistic ion transport models at the nano-scale and optimal process design through deterministic global superstructure optimization at a large-scale. Thereby, we facilitate ANN surrogate models trained on data generated by a one-dimensional ion transport model, which are subsequently embedded in a state-of-the-art membrane process optimization model. 

We use the extended Nernst-Planck model, called pEnPEn, for describing the ion transport through the membrane~\cite{femmer2015ion, femmer2016mechanistic, evdochenko2019direct}. The model pEnPEn describes ion transport through multi-layered geometries (i.e., LbL nanofiltration membranes) consisting of $n$ electrolyte layers (En) with $n$ polyelectrolyte layers, i.e., membranes, (PEn). An applied pressure ($p$) acts as the driving force of the separation process. 
The aforementioned one-dimensional ion transport model is evaluated offline, creating an extensive data set that describes the membrane's ionic retention based on membrane-specific parameters (such as, the layer charge $X$ and the layer thickness $\Delta x$) and process-specific parameters (such as, the transmembrane velocity $v$, and ionic feed concentration $c_j$). A subsequent training of the data creates an ANN surrogate model. 

\pagebreak
Next, the surrogate models are exploited towards a more accurate two-dimensional distribution of the membrane module even to capture the filtration-related decreasing retention of salt. Therefore, individual ANNs are arranged in series to resolve the membrane in the direction of flow. The data that is generated by this series of ANNs are used to create a two-dimensional surrogate model, which is then embedded in the membrane process optimization model. 
This enables the complex mechanistic model to be applied computational resource-efficiently in an optimization context, despite the partial differential-algebraic form of the original model.

The paper is organized as follows: 
First, we present the data generation and the learning of ANN surrogate models. Second, the more accurate two-dimensional distribution of the membrane module is presented through ANN surrogate modeling. Third, the hybrid mechanistic / data-driven multi-scale process model is applied to a case study optimizing nanofiltration membrane plants for multiple objectives, feed concentrations, filtration stages, and salt mixtures. 
The membrane module synthesis properties are optimized along with the superstructure of the membrane plant. In particular, process configurations of the plant (i.e., number of filtration stages, recirculating of streams, pump power) as well as the membrane synthesis properties (i.e., the layer charge $X$ and the layer thickness $\Delta x$) are degrees of freedom and are optimized simultaneously. 

This work sets the foundation for computational resource-efficient multi-scale modeling integrating simulation studies of the complex ion transport in the decision making process at large process scale optimizations.
Finally, all developed models and the optimization solvers are available open-source, making it a viable tool for future research and industrial applications.
%
%
%
%
%
%
%
%
%
%
\newpage
\section{Methodology}

In this section, the workflow of the data-driven approach is described to bridge the scales between (i) high fidelity ion transport models on a nano-scale and (ii) deterministic global membrane process optimization on a large-scale. 
First, the high fidelity simulation of the ion transport through the membrane on the nano-scale is replaced by a surrogate model. This surrogate model based replacement is performed by evaluating the high fidelity ion transport model offline and creating a training data set that is subsequently learned by a supervised machine learning algorithm (here ANNs). 
Then, the resulting data-driven ANN-based surrogate model that approximates the accurate model is embedded in a hybrid mechanistic / data-driven model and optimized to find optimal membrane plant process configurations.
In summary, this procedure avoids PDE constraints in the formulation of the deterministic global optimization problem. Thereby, this method enables deterministic optimization of membrane processes with surrogate models at a low computational cost similar to lumped systems (i.e., heuristics or short cut models) but significantly more accurate.

\begin{figure}[h]
  \centering
    \includegraphics[width=1\linewidth]{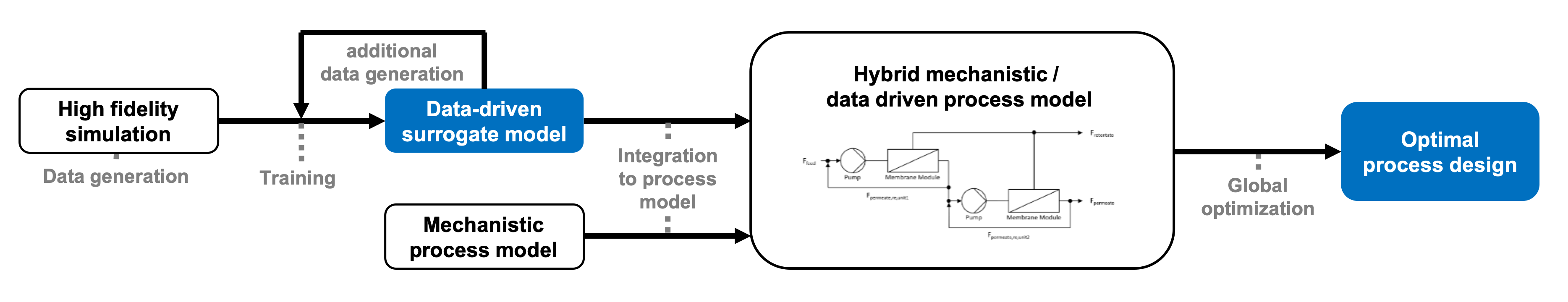}
        \vspace{-1cm}
        \caption{The workflow of the data-driven approach. The input data is created through the high fidelity ion transport model simulation study. Next, an ANN takes this data to learn the ionic separation performance of the membrane based on the model, creating a data-driven ANN-based surrogate model. Two-dimensional distribution of the membrane module is constructed for higher fidelity of the surrogate model. For this purpose, the surrogate model is used to generate additional data that is trained in a new ANN surrogate model. Next, this ANN surrogate model is embedded in a mechanistic process model. Finally, the optimal process design is solved by deterministic global process optimization.}
        \label{fig:DataDrivenWorkflow}
\end{figure}

The workflow commences as depicted in Figure~\ref{fig:DataDrivenWorkflow}. 
(1) The high fidelity ion transport model (Section~\ref{sec:datageneration}) is evaluated offline to create an extensive data set, that (2) is used to train the data-driven ANN-based surrogate models on the data (Section~\ref{sec:ANNsMethods}). 
(3) Moreover, a surrogate model, including a two-dimensional distribution of the membrane module is created for higher fidelity. The most precise two-dimensional distribution and additional data generation are described in Section~\ref{sec:ANNsMethodsadaption}, and the implementation of the two-dimensional surrogate model is described in Section~\ref{sec:ANNsTwoSurrogate}.
(4) Next, the individual surrogate models are integrated with a mechanistic process model to a hybrid mechanistic / data-driven process model. In Section~\ref{sec:membrane_process_design}, we explain the overall mechanistic process model, including cost correlations. 
(5) Finally, the hybrid model is optimized using a reduced-space formulation~\cite{schweidtmann2019deterministic} and our open-source deterministic global solver MAiNGO~\cite{MAiNGO}. In Section~\ref{sec:global_deterministic_optimization}, we briefly describe the optimization method.

All tools necessary to rebuild and extend this method are open-source available to everyone. The process models and optimization problems are available open-source (\url{http://permalink.avt.rwth-aachen.de/?id=506639}). The training data, the trained artificial neural networks, and the optimization results are available in the electronic supplementary material of this article (published by the journal). The artificial neural network models and training scripts are available open-source within the ``MeLOn - \textbf{M}achin\textbf{e} \textbf{L}earning Models for \textbf{O}ptimizatio\textbf{n}'' toolbox~\cite{MeLOn_Git} (\url{https://git.rwth-aachen.de/avt.svt/public/MeLOn/}). Furthermore, the deterministic global solver MAiNGO is also available open-source (\url{https://git.rwth-aachen.de/avt.svt/public/maingo}). 
%
%
%
%
%
%
%
%
%
%
\subsection{Data generation by high fidelity ion transport model}\label{sec:datageneration}

\begin{figure}[h]
  \centering
  \includegraphics[width=0.75\linewidth]{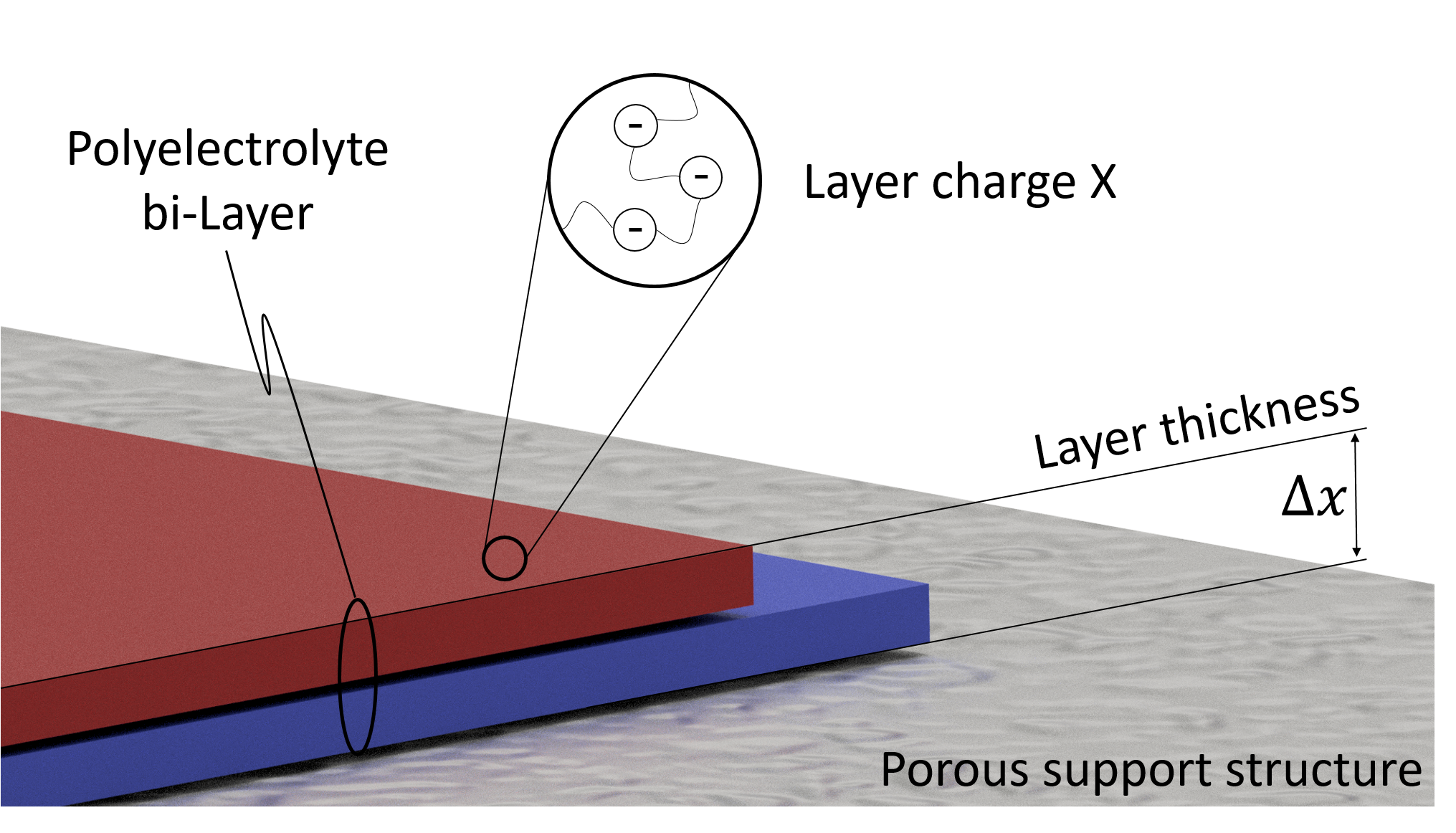}
        \vspace{-0.5cm}
		\caption{Overview of model parameters of the extended ion transport model pEnPEn~\cite{femmer2015ion, femmer2016mechanistic, evdochenko2019direct}. The layer charge $X$, and the layer thickness $\Delta x$ of the separation layer determine the separation performance. These values are inputs to the ANN and serve as a degree of freedom during simultaneous optimization of membrane module synthesis properties along with the superstructure of the membrane plant.}
		\label{fig:MechanisticModelIonicTransport}
\end{figure}

In this work, the initial data set to train the ANN surrogate model is created by a simulation study using a high fidelity mechanistic ion transport model. We use a one-dimensional extended Nernst-Planck model, called pEnPEn~\cite{femmer2015ion, femmer2016mechanistic}, to describe the ion transport through multi-layered geometries. LbL nanofiltration membranes can be modeled with this model~\cite{evdochenko2019direct} by introducing the layer thickness $\Delta x$ of the separation layer with a homogeneous layer charge $X$ as shown in Figure~\ref{fig:MechanisticModelIonicTransport}. 

The model is evaluated offline, creating a large data set that describes the membrane's ionic retention based on membrane-specific parameters (such as, the layer charge $X$ and the layer thickness $\Delta x$) and process-specific parameters (such as, the transmembrane velocity $v$, ionic feed concentration $c_j$). The generation of data is automated, utilizing a wrapper script. This generates a set of training data inputs/outputs by a Latin hypercube sampling design for the ANN surrogate model. The variables are listed as follows: 
\begin{itemize}
    \item The \textbf{layer charge $X$~[\SI{}{\mol\per\cubic\meter}]} is a membrane-specific variable and \textbf{input} to the ANN. For single salts the layer charge varies in the interval $[-500, -100] \subset \mathbb{R}$. For salt mixtures this interval needs to be extended to $[-1000, -10] \subset \mathbb{R}$ to account for the more challenging separation task. The membrane charge needs to be higher as compared to single salts due to the consideration of higher ionic concentrations for salts mixtures. 
    \item The separation \textbf{layer thickness $\Delta x$~[\SI{}{\nano\meter}]} is a membrane-specific variable and \textbf{input} to the ANN. The separation layer thickness varies in the interval $\in [75, 150] \subset \mathbb{R}$ according to experimental observations of \cite{menne2016precise}.
    \item The \textbf{transmembrane velocity $v$ [\SI{}{\micro\meter\per\second}]} is a process-specific variable and \textbf{input} to the ANN. The retention of ions strongly depends on the transmembrane velocity, i.e., recovery rate, and varies in this study in the interval $[0, 50] \subset \mathbb{R}$.
    \item The \textbf{salt feed concentration $c_j$~[\SI{}{\mol\per\cubic\meter}]} is a process-specific variable and \textbf{input} to the ANN. The retention of ions strongly depends on the feed concentration. The range of salt feed concentration for a single membrane unit is $[1, 50] \subset \mathbb{R}$. When using two membrane filtration units in series, an extension of the data for feed concentrations below \SI{1}{\mol\per\cubic\meter} is needed. Therefore, the data is extended using an additional Latin hypercube sampling to extend the data towards concentrations in the interval $[0.01, 50] \subset \mathbb{R}$.
    \item The \textbf{ionic retention of ion j $R_j$ [\%]} is the \textbf{output} to the ANN.
\end{itemize}

The one-dimensional model is established for the membrane itself, along with two adjacent diffusive layers. The model is implemented in COMSOL Multiphysics\textsuperscript{\textregistered}~\cite{COMSOL} and executed via the MATLAB~\cite{MATLAB2019} API. 
The partial differential extended Nernst-Planck equation is simplified to a second-order ordinary differential equation by considering only the steady-state behavior of the charged species. Moreover, the pressure-driven convective flux is described via the first-order ordinary differential equation Darcy’s law. The boundary conditions are applied for the pressure at the inlet and outlet of the domain. Additionally, charge neutrality across the membrane is assumed. 
The resulting problem formulation involves the coupling of two different ordinary differential equations and is highly non-linear. It is solved with an adaptive Newton-Rapson algorithm. The systems matrix of the linearized problem is inverted using the MUMPS algorithm. This modeling approach is generic and thus applies to all combinations of salts used in this study.

Due to convergence issues in the range of low initial salt feed concentrations (i.e., at high local gradients in the second-order ordinary differential equation), a load ramping technique is applied to the excess charge parameter on the membrane surface. Input combinations to the one-dimensional ion transport model for which the solver did not converge are excluded from the training set. 
%
%
%
%
%
%
%
%
%
%
\subsection{Training of the data-driven surrogate model}\label{sec:ANNsMethods}

In this study, we utilize supervised machine learning techniques for the creation of surrogate models. In literature, there exists a large variety of machine learning techniques, including Gaussian processes, artificial neural networks, and decision trees for correlations of varying complexity. In this work, we implement ANNs as surrogate models, as they can capture the ion transport in membranes based on synthesis parameters, as demonstrated in our previous work~\cite{rall2019rational, rall2020simultaneous}. ANNs are black box models that are frequently used for unsupervised and supervised learning tasks, including pattern recognition, classification, and regression~\cite{lecun2015deep}. ANNs learn training data and can approximate an underlying input-output function. The evaluation of ANNs is given by an explicit algebraic function that depends on the architecture of the ANN, i.e., the connection of neurons~\cite{dayhoff2001artificial}. This function is embedded in the process optimization problem and is efficiently solved by a reduced-space deterministic global optimization method~\cite{schweidtmann2019deterministic}.

\begin{figure}[H]
  \centering
  \includegraphics[width=1\linewidth]{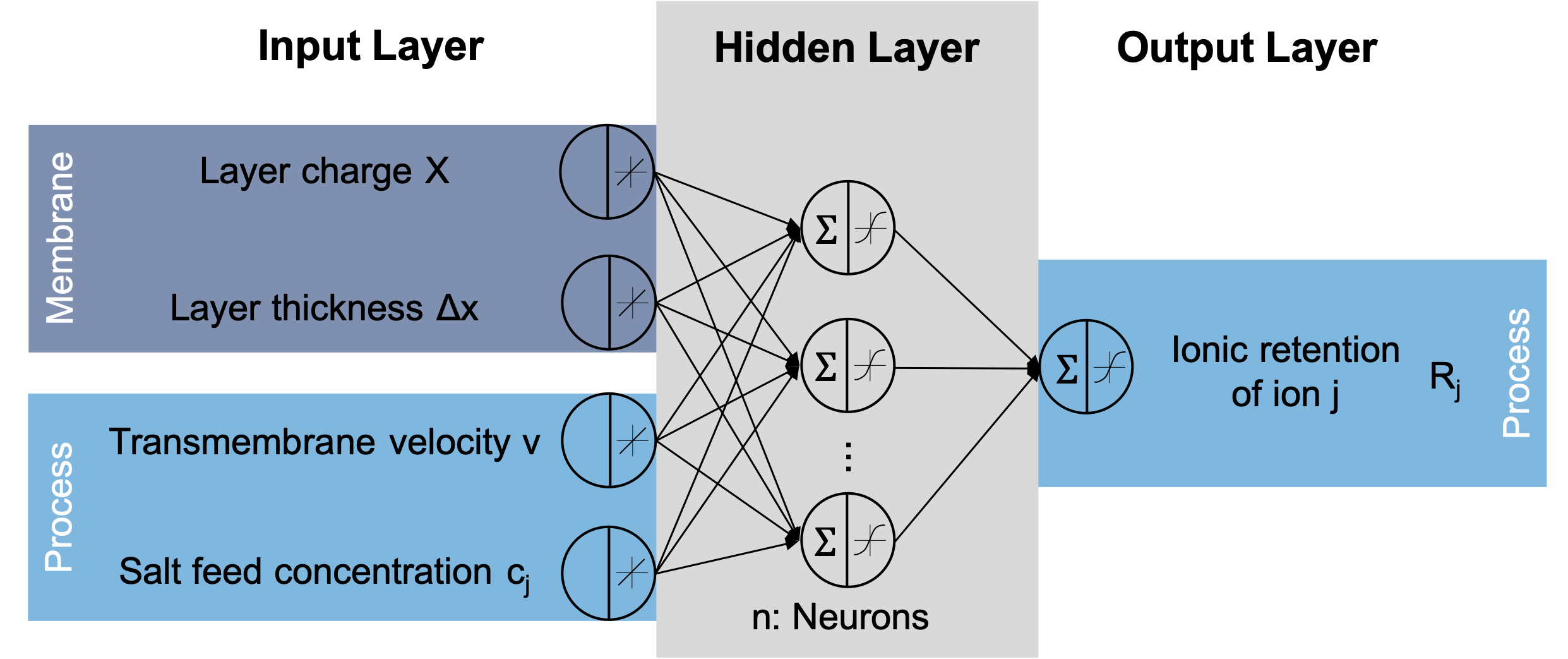}
        \caption{Multi-layered perceptron ANN with one hidden layer used this study as a data-driven surrogate model to recreate the ion transport through nanofiltration membranes. The training is based on data generated by a one-dimensional extended Nernst-Planck mechanistic ion transport model (pEnPEn). The membrane-specific parameters (such as the layer charge $X$ and the layer thickness $\Delta x$) and process-specific parameters (such as the transmembrane velocity $v$ and salt feed concentration $c_j$) are inputs to the ANN. The ionic retention of ion j $R_j$ is the output to the ANN.}
        \label{fig:ANNiontransport}
\end{figure}

In this work, a shallow feed-forward multi-layered perceptron ANN including one hidden layer is utilized to recreate the one-dimensional extended Nernst-Planck mechanistic ion transport model (pEnPEn) from the previous section. A hyperbolic tangent transfer function is employed in the hidden- and output layer. The structure of the ANN is depicted in Figure~\ref{fig:ANNiontransport}. Inputs to the ANN are the layer charge $X$, the layer thickness $\Delta x$, the transmembrane velocity $v$, and the salt feed concentration $c_j$. Output to the ANN is the ionic retention of ion j $R_j$. 

The main drawback of the data-driven modeling approach using ANNs is its limited extrapolation capability. Meaning that any data-driven model is not valid anymore outside of its training data domain (i.e., its convex envelope). Therefore, this results in a large data requirement for the training of ANNs. In our study, the training data of the ANNs is generated by a mechanistic ion transport model that allows us to have any number of data points (within computational reason) for training and validation. A consideration of probabilities, i.e., through Gaussian processes, is omitted because we use a data set extracted from a mechanistic model.
The training data set we use for this study is small compared to many other machine learning applications, such as deep artificial neural networks with millions of parameters that are frequently used for image recognition. Thus, we select a shallow artificial neural network architecture with one hidden layer ranging from 6 to 14 hidden neurons because the input dimensionality and training data set are much smaller. Furthermore, the training of the ANNs is optimized to minimize adverse effects using small data sets including: (a) multiple runs of ANN trainings to mitigate sporadic fluctuations in artificial neural network performance, (b) a randomized data set for training, validation and test to account for random effects due to small test data, and (c) the use of a k-fold cross-validation to prevent overshooting between data sets and avoid overfitting of the data set.

The data preparation and training procedure are based on our previous work~\cite{rall2019rational, rall2020simultaneous}, and adapted accordingly. The architecture of ANNs is flexible and can be adjusted by setting its hyperparameters, e.g., number of layers, for application. In general, the architecture should be selected appropriately to avoid undesired under- and overfitting effects. The available data set is split into 70~\% training, 15~\% validation, and 15~\% test set at random. Early stopping is used to prevent overshooting during all ANN training. All ANNs used in this work are created in the MATLAB environment~\cite{MATLAB2019,marquardt1963algorithm}. The training procedure commences in the following steps: 
First, the hidden layer size is determined by k-fold crossvalidation with a range of $n \in [1, 50] \subset \mathbb{N}$ hidden neurons. The minimum of the characteristic bias/variance trade-off bathtub curve of the mean-squared-error (MSE) determines the best number of hidden neurons $n$ and thereby the hidden layer size.
Second, ANNs are trained manifold with the previously determined hidden layer size $n \pm 3$ using a split of 70~\% training, 15~\% validation, and 15~\% test set at random.
Third, the best performing candidate of the ANN is selected for the following case studies. 

Two ANNs are utilized as one dimensional data-driven surrogate model considering a single salt \ce{Na2SO4} or either a salt mixture of \ce{NaCl} and \ce{Na2SO4}. Here, only the predominant ion species (i.e., the ions \ce{Cl-} and \ce{SO4^{2-}}) of the filtrated salt mixture are trained as inputs to the ANN. This consideration is valid due to electroneutrality in the solution and membrane. The monovalent cation \ce{Na+} is calculated based on a mass balance. For single salts the ANN has one output (i.e., \ce{Na2SO4} retention) and for salt mixtures, the ANN has two outputs (i.e., \ce{Cl-} retention and \ce{SO4^{2-}} retention). Negative salt retentions are possible \cite{yaroshchuk2008negative} and are mapped by the ANN. A list of properties of both ANNs is summarized in Table~\ref{tab:ArificialNeuralNetworks}. 

\begin{table}[h]
    \caption{One-dimensional data-driven surrogate models used in this work to recreate the one-dimensional extended Nernst-Planck high fidelity mechanistic ion transport model (pEnPEn). The number of neurons $n$ of each ANN, the amount of training data set entries, and the section in which it is used is given.}
	\centering
	\label{tab:ArificialNeuralNetworks}
    \begin{tabular}{lclrc}
        \hline
                                   Network name &  Hidden neurons $n$   & Description   & Training points      & Used in Section          \\
        \hline
        ANN$_{Na_{2}SO_{4},~k=1,~ 1~\text{Unit}}$             &  6        & Single salt   & 998                  & \ref{sec:ANNsMethodsadaption}, \ref{sec:results_optimal_solution_strategy}    \\
        ANN$_{NaCl,~Na_{2}SO_{4},~k=1,~1~\text{Unit}}$         &  8        & Salt mixture  & 1,000                 & \ref{sec:ANNsMethodsadaption}    \\
        \hline
    \end{tabular}
\end{table}
%
%
%
%
%
%
%
%
%
%
\subsection{Two-dimensional distribution of the membrane module}\label{sec:ANNsMethodsadaption}

Up until now, the membrane unit is considered where only the changes in the orthogonal direction of the membrane are considered by the one-dimensional surrogate model of the previous section. However, due to the retention of salt at the membrane, the feed concentration in the module increases with the length of the membrane. The feed concentration of the salt strongly influences the retention of the nanofiltration membrane. Changes in feed concentration in the direction of flow are not considered so far. 

\begin{figure}[H]
  \centering
  \includegraphics[width=.95\linewidth]{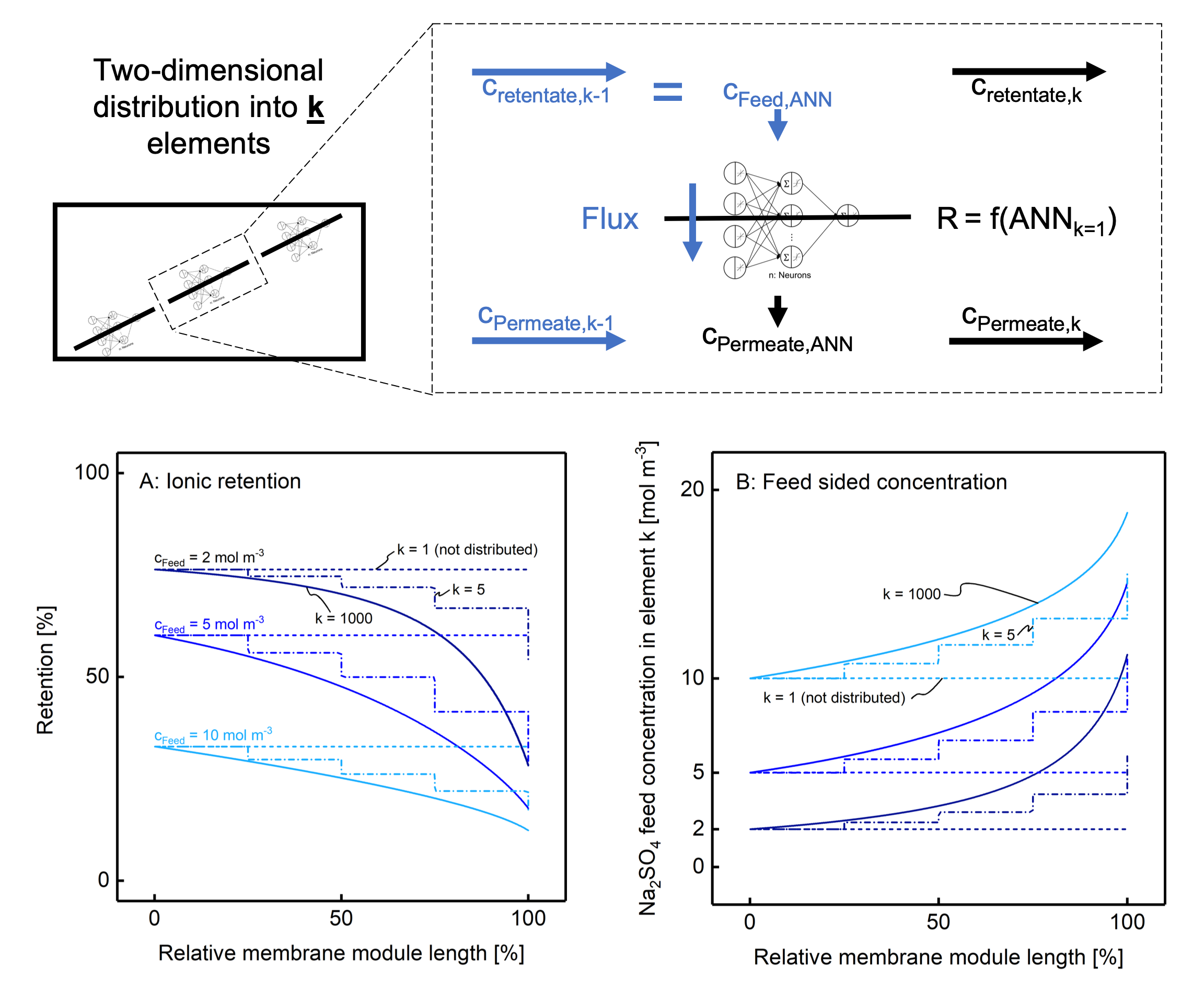}
		\caption{Two-dimensional distribution of the membrane module in flow direction through ANN-based surrogate models. Shown is the negative effect of the feed concentration on the salt retention of the membrane. Only the two-dimensional distributed cases do not underestimate this local reduction of ionic retention in the direction of flow. In A), the retention of one selected case and membrane are shown for three different initial feed concentrations and varying element numbers $k$. In b), the corresponding feed sided concentration is shown resulting from the filtration-related increase of feed concentration.}
		\label{fig:results_module}
\end{figure}

In Figure~\ref{fig:results_module}, a two-dimensional distribution of the membrane module is proposed in flow direction through ANN-based surrogate models. Here, the membrane length is divided into several elements $k$ and solved for the conservation equations on every element. An individual ANN describes each element. In Figure~\ref{fig:results_module}~A-B the strong influence of the feed concentration of the salt on the retention is evident for a selected membrane case. 

Two effects contribute to the reduction of salt retention. First, the retention decreases with increasing initial salt feed concentration. This is demonstrated in Figure~\ref{fig:results_module}~A-B for the concentrations of \SI{2}{\mol\per\cubic\meter}, \SI{5}{\mol\per\cubic\meter}, and \SI{10}{\mol\per\cubic\meter}. Second, a filtration-related reduction of the retention of salt is observed in the direction of flow. During filtration, the salt concentration increases at the retentate side. This negative effect of the increasing feed concentration on the salt retention of the membrane is differently pronounced for the shown three cases. A one-dimensional distributed membrane underestimates this effect and leads to inaccurate results. With an increasing number of elements $k$, the approximation approaches the most accurate concentration profile. The relative gap between a one-dimensional distributed membrane and a two-dimensional distributed membrane is calculated for the specific case in Table~\ref{tab:results_RelativeGapDiscretization}. Here, the relative gap ranges from +43\% to +240\% dependent on the initial feed concentration. These results are case-specific but apply to any other case. 

\begin{table*}[h]
	\caption{The Feed concentration increases along the membrane length during filtration. The relative gap of the permeate concentration is calculated based on a two-dimensional distributed membrane with $k=1000$~elements and a one-dimensional distributed membrane with $k=1$~element for different feed concentrations. The gab occurs due to the decreasing retention with increasing feed concentration which is not accounted for in a one-dimensional distributed membrane.}
	\centering
	\label{tab:results_RelativeGapDiscretization}
	\begin{tabular}{rcccc}
	\hline
   Variable name    & Unit                              & $c_{feed}$ = \SI{10}{\mol\per\cubic\meter}       & $c_{feed}$ = \SI{5}{\mol\per\cubic\meter}        & $c_{feed}$ = \SI{2}{\mol\per\cubic\meter}        \\
      \hline
Two-dimensional distributed ($k = 1000$)  & [\SI{}{\mol\per\cubic\meter}]     & 6.71                                            & 1.99                                            &  0.47                                          \\
One-dimensional distributed ($k = 1$) & [\SI{}{\mol\per\cubic\meter}]     & 9.62                                            & 4.58                                            &  1.61                                          \\
\hline
Absolute difference            & [\SI{}{\mol\per\cubic\meter}]     & 2.92                                            & 2.53                                            &  1.14                                          \\
Relative difference            & [\SI{}{\percent}]                 & +43.50                                          & +129.75                                         &  +239.86                                       \\
\hline
\end{tabular}
\end{table*}

\begin{figure}[H]
  \centering
  \includegraphics[width=0.6\linewidth]{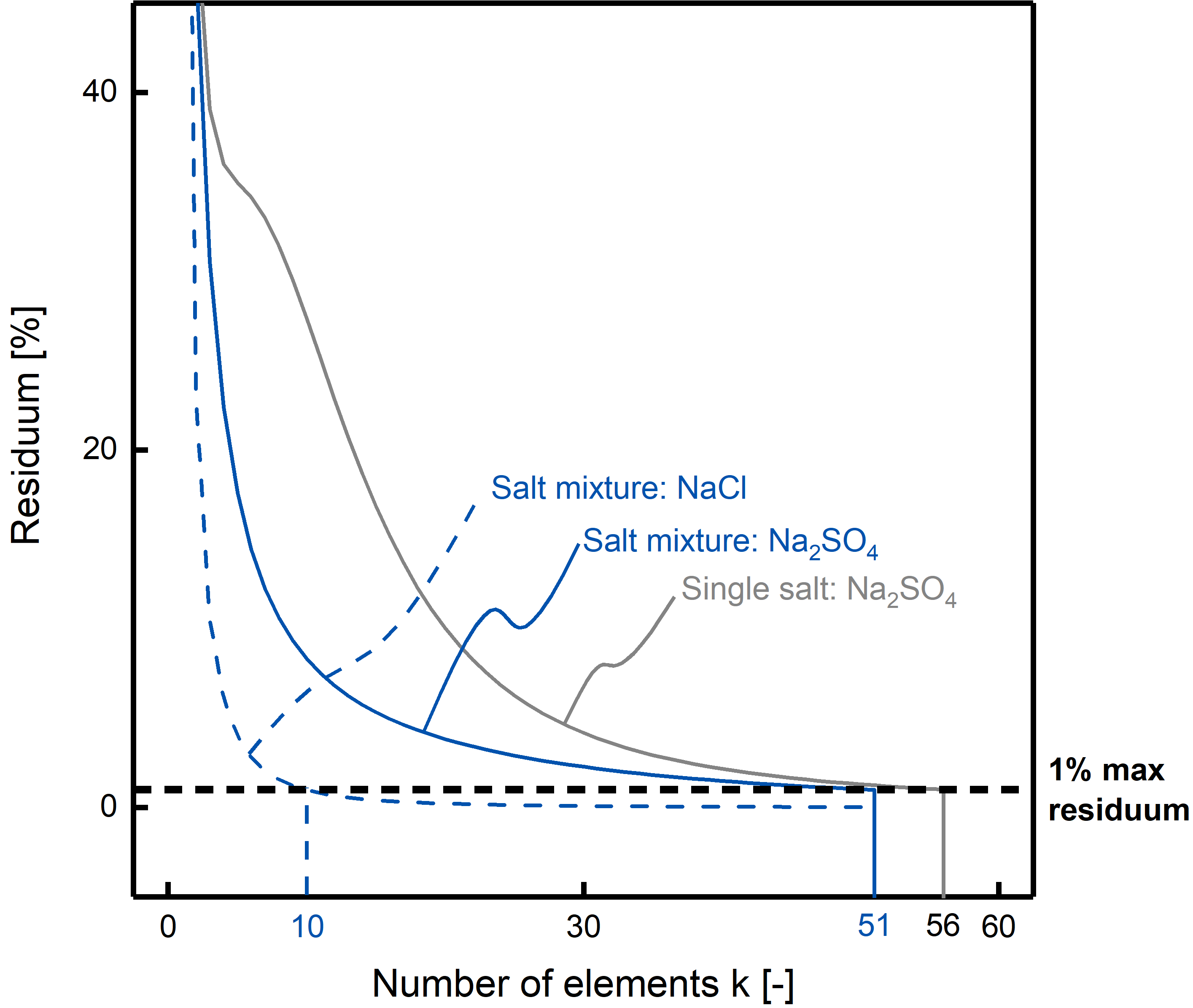}
        \caption{Two-dimensional distribution of the module in flow direction: The membrane unit is split into multiple elements $k$. The number of elements is increased until the relative gap of filtration performance through the permeate concentration is below 1\%.}
        \label{fig:results_ResiduumVSElements}
\end{figure}

Next, the most accurate number of elements for a single salt or salt mixture process optimization is identified. This is necessary to determine the number of two-dimensional distribution elements $k$ that are necessary to describe the change of states along the membrane surface accurately. Therefore, the number of elements is increased until the relative gap of the concentration is below 1\%, where it can be considered sufficiently accurate. The relative difference (or residuum) is calculated by the relative gap between the two-dimensional distributed ($k>1$), and one-dimensional distributed ($k=1$) permeate concentration leaving the membrane module. This procedure is repeated for different membrane configurations and observed covering the whole application range. Two exemplary cases and the progression of the residuum over the number of elements used are visualized in Figure~\ref{fig:results_ResiduumVSElements}. Here, a maximum number of $k=56$ elements are needed for single salt applications, and $k=51$ elements are required for salt mixtures. 

%
%
%
%
%
%
%
%
%
%
\subsection{Training of the surrogate model with two-dimensional distribution of the membrane}\label{sec:ANNsTwoSurrogate}

Resolving the membranes through two-dimensional distribution is achieved by subdividing it into a fixed number of elements and arranging the individual ANNs in series. Then, the mass conservation laws and the ANN computations are employed on every element. This creates a local output state from the input for each element. The technique is recursively applied to all elements along the direction of flow.

Now, this repeated computation of the local element quantities can be directly embedded into the process of plant optimization. However, this nesting of ANNs can lead to weak McCormick relaxations and can adversely affect the computational performance of global optimization~\cite{Doncevic2020GlobalANNControl}. 
An alternative approach is to train a new ANN surrogate model that includes these recursions as a black box creating a single surrogate model for the two-dimensional modeling of the nanofiltration module. The training of an ANN-based surrogate model, including a two-dimensional distribution of the membrane module, is described in the following.

A special case of data generation and ANN training arises when creating the surrogate model, including a two-dimensional distribution. The one-dimensional surrogate models trained on the one-dimensional extended Nernst-Planck ion transport model are connected serially one after the other and evaluated offline. This series of surrogate models create an additional data set for the new two-dimensional surrogate model. This extended two-dimensional surrogate model contains additional influential variables as the volume flows of the entire module need to be considered. A partial resolution of the recursive equation (exemplary for the single salt case) results in the following description

\begin{align}
    c_{in}(k) = c_{in}(k-1) \cdot \left( 1 + \frac{\gamma}{K-(k-1) \cdot \gamma} R(k-1) \right)
    \label{eqn:discr1}
\end{align}

where $c_{in,j}$ denotes the feed concentration in element $k$, $K$ the total number of elements, $R_j$ the ionic retention, and $\gamma$ the flow ratio which denotes the ratio of feed and permeate flow along with the nanofiltration unit. Where $\gamma$ is defined as 

\begin{align}
    \gamma = \frac{F_{\text{permeate}}}{F_{\text{feed}}}.
    \label{eqn:discr_kappa}
\end{align}

This flow ratio of the membrane module determines the performance of the unit. Therefore, the two-dimensional surrogate model is extended by the additional input of $\gamma$ to account for the extra parameter.

For obtaining the training data of the two-dimensional surrogate model, an additional Latin hypercube sampling for a random two-dimensional distribution is applied, and the data created offline. Due to the low computational effort to calculate such recursive ANNs in series, $k = 1000$ elements is chosen for the following process optimizations. Therefore, the relative difference between the two-dimensional distributed ($k>1$) and one-dimensional distributed ($k=1$) permeate concentration leaving the membrane module is always below 1\%. For example, in Figure~\ref{fig:results_ResiduumVSElements} only a maximum number of $k=56$ elements is needed. 

The training procedure is performed as previously stated in Section~\ref{sec:ANNsMethods}. Now, inputs to the ANN are the layer charge $X$, the layer thickness $\Delta x$, the transmembrane velocity $v$, the salt feed concentration $c_j$, and the additional parameter $\gamma$ for the flow ratio. The flow ratio $\gamma$ is valid in the range $\in [0.18; 1] \subset \mathbb{R}$ under the consideration of the intercorrelation of $\gamma$ (i.e., the permeate flow can never be higher than the feed flow). Output to the ANN is the ionic retention of ion j $R_j$. A list of properties of the new two-dimensional surrogate models can be found in Table~\ref{tab:ArificialNeuralNetworks2}. 

\begin{table}[h]
    \caption{Two-dimensional surrogate models used in this work to recreate the two-dimensional distribution of the membrane in the flow direction. The number of neurons $n$ of each ANN, the amount of training data set entries, and the section in which it is used is given.}
	\centering
	\label{tab:ArificialNeuralNetworks2}
    \begin{tabular}{lclrc}
        \hline
                                   Network name &  Hidden neurons $n$   & Description   & Training points      & Used in Section          \\
        \hline
        ANN$_{Na_{2}SO_{4},~k=1000,~1~\text{Unit}}$           &  6        & Single salt   & 998                  & \ref{sec:results_optimal_solution_strategy}, \ref{sec:results_single_salt}    \\
        ANN$_{Na_{2}SO_{4},~k=1000,~2~\text{Units}}$          &  6        & Single salt   & 1,198                 & \ref{sec:results_single_salt}    \\            
        ANN$_{NaCl,~Na_{2}SO_{4},~k=1000,~1~\text{Unit}}$      &  14       & Salt mixture  & 1,000                 & \ref{sec:results_salt_mixture}    \\    
        \hline
    \end{tabular}
\end{table}
%
%
%
%
%
%
%
%
%
%
\subsection{Hybrid mechanistic / data-driven process superstructure model}\label{sec:membrane_process_design}

\begin{figure}[H]
  \centering
  \includegraphics[width=0.75\linewidth]{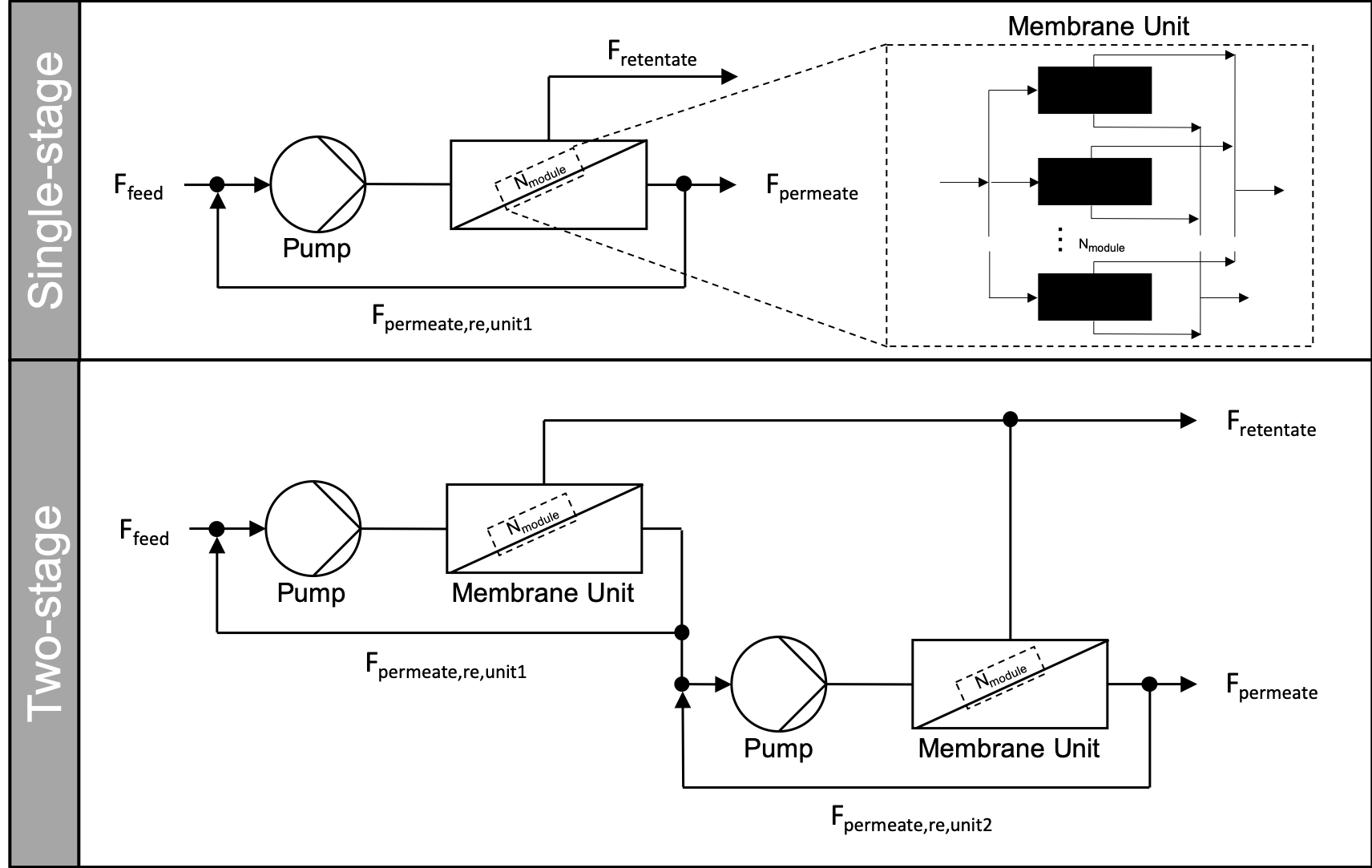}
		\caption{Layouts of the membrane process plant consisting of either a single nanofiltration unit or a two-stage connection of nanofiltration units. Permeate flow recirculation is allowed but does not have to be used. Each of the membrane units is equipped with a feed sided pump. A global feed, and retentate and permeate are connected by mixing units and represent the process input and outputs}
		\label{fig:Superstructure}
\end{figure}

The individual surrogate models are integrated with a mechanistic process model to a hybrid mechanistic / data-driven process model. The hybrid process superstructure model comprises of a mechanistic process model to describe the process plant and of surrogate models to describe the ionic transport based on membrane and process-specific variables. The mechanistic process model consists of component mass balances, a pump model, and cost correlations. The superstructure of the process plant is shown in Figure~\ref{fig:Superstructure}. The model is adapted from our previous work~\cite{rall2020simultaneous}. 

Either a single-stage nanofiltration unit or a two-stage connection of nanofiltration units is chosen for the process plant. A pump is installed in the inlet stream of each nanofiltration unit to provide the main driving force. Each nanofiltration unit consists of multiple membrane modules, $N_{module}$ in parallel. Each nanofiltration unit~(unit $i$) has a feed stream $F_{feed,i}$. A concentrated retentate stream $F_{retentate,i}$ and a diluted permeate stream $F_{permeate,i}$ each leave the nanofiltration unit. The permeate stream $F_{permeate,i}$ is calculated based on the membrane's ionic retention performance, permeability, and pressure difference $\Delta p$ applied:
\begin{align}
	F_{permeate,i} = Q \cdot A_{unit,i} \cdot \Delta p
	\label{eqn:cross}
\end{align}
where $Q$ is the permeability of the membrane unit, $A_{unit,i} = N_{module,i} \cdot A_{module}$ is the total membrane area of the membrane unit. The size of a single membrane module in a nanofiltration unit is set to $A_{module}$ = \SI{60}{\square\meter} in accordance with multiple membrane modules available on the market~\cite{pentair, dupontUF}. The permeability of the membrane is estimated based on a linear correlation of the membrane thickness $\Delta x$ with the range $Q \in [5; 35] \subset \mathbb{R}$ in accordance with laboratory results obtained in our previous work~\cite{menne2016precise}. 
\begin{align}
	Q = 0.4 \cdot \Delta x - 25
	\label{eqn:Q=f(delta)}
\end{align}

Additionally, the layer charge $X$ is assumed to increase linearly with membrane thickness $\Delta x$. A mass balance coupled with the ionic retention of ion j $R_{i,j}$ for ion $j$ closes the governing equations. 

\begin{align}
	F_{feed,i}\hspace{0.1cm}c_{feed,i,j} = F_{retentate,i}\hspace{0.1cm}c_{retentate,i,j} + F_{permeate,i}\hspace{0.1cm}c_{permeate,i,j}
	\label{eqn:balance}
\end{align}
\begin{align}
	R_{i,j} = 1 - \frac{c_{permeate,i,j}}{c_{feed,i,j}}
	\label{eqn:ret}
\end{align}

The ionic retention of ion j $R_{i,j}$ is computed by the ANNs as a function of membrane-specific variables and process-specific variables as described in Section~\ref{sec:ANNsMethods} and Section~\ref{sec:ANNsTwoSurrogate}.

The cost model for this study is adapted from our previous work~\cite{rall2020simultaneous}. Here, the Verberne~cost~model~\cite{verberne1993membraanfiltratie, sethi2000cost, ang2017effect} is used. The total investment cost $C_{investment}$ is calculated based on the sum of the civil $C_{civil}$, mechanical $C_{mechanical}$, electromechanical $C_{electro}$, and membrane $C_{membrane}$ investment cost
\begin{align}
	C_{investment} = C_{civil} + C_{mechanical} + C_{electro} + C_{membrane}.
\end{align}

The annual operation cost is calculated based on the sum of the depreciation cost $C_{depreciation}$, the energy cost $C_{energy}$, the maintenance cost $C_{maintenance}$, and specific cost $C_{specific}$.
\begin{align}
	C_{operation} = C_{depreciation} + C_{energy} + C_{maintenance} + C_{specific}
\end{align}

Appropriate cost parameters and depreciation periods are accounted for by the individual investment cost~\cite{rall2020simultaneous}.

\vspace{0.5cm}
This study does not consider the effects of osmotic pressure differences on the reduction of the driving force (here transmembrane pressure). This reduction of the driving force means that the presented results underestimate the pumping power because of the high salt concentrations used in this study. Considering the previously described cost model, it might be more economically viable to implement more membrane modules per membrane unit than using a higher pressurized feed stream. There always exists a trade-off between the costs by increased pumping power and the capital costs of more membrane area. Thus, the integration of the osmotic pressure into the proposed framework would be relevant and important for future research. In particular, future work could consider osmotic pressure differences in the ion transport models by accounting for the osmotic pressure in the two-dimensional model used for data generation (c.f. Section~\ref{sec:ANNsMethodsadaption}-\ref{sec:ANNsTwoSurrogate}). However, this is beyond the scope of the current study. Considering the effects of osmotic pressure differences on the reduction of the driving force demand for a more detailed investigation as well as validation of the model against existing methods.
%
%
%
%
%
%
%
%
%
%
\subsection{Numerical optimization approach}\label{sec:global_deterministic_optimization}

Finally, the hybrid model is optimized using a reduced-space formulation and our open-source deterministic global solver MAiNGO~\cite{MAiNGO}. The optimization problems in this work are formulated as a multi-objective mixed-integer nonlinear program (MINLP):
\begin{align}\label{eqn:MINLP}
\begin{split}
\min_{{\bf x},{\bf y}} \quad & \begin{pmatrix}f_1(\textbf{x},\textbf{y}) \\ f_2(\textbf{x},\textbf{y})\end{pmatrix} \\
s.t. \quad &g_{i}(\textbf{x},\textbf{y}) = 0, \quad i=1,...,I \\
&h_{j}(\textbf{x},\textbf{y}) \leq 0, \quad j=1,...,J \\
&\textbf{x} \in X \subset \mathbb{R}^{n}, \quad
 \textbf{y} \in Y \subset \mathbb{Z}^{m} \\ 
\end{split}
\end{align}
where $g_{i}(\textbf{x},\textbf{y})$ are equality constraints, $h_{j}(\textbf{x},\textbf{y})$ are inequality constraints,
$f_1(\textbf{x},\textbf{y}),f_2(\textbf{x},\textbf{y})$ are the objective functions, $\textbf{x}$ are continuous optimization variables, and $\textbf{y}$ is the integer optimization variable (i.e., number of membrane modules $N_{module,i}$). 
All model equations for the mechanistic process model are described in Subsection~\ref{sec:membrane_process_design}. The ionic separation performance is mapped by ANNs as described in Subsection~\ref{sec:datageneration}, which are formulated in a reduced-space formulation~\cite{schweidtmann2019deterministic}. The objective functions are (i) minimal annual operation costs ($f_1(\textbf{x},\textbf{y}) = C_{operation}$) and (ii) minimal permeate concentration ($f_2(\textbf{x},\textbf{y})=c_{permeate}$).

The MINLP is solved for the two objectives by the $\epsilon$-constraint method. Here, the multi-objective problem is reformulated to multiple single-objective problems. Thereby, one objective is minimized, and the other objective is enforced to be less or equal to a parameter $\epsilon$. This procedure is repeated for different $\epsilon$ yielding Pareto-optimal points. All optimization problems in this work are implemented and solved by our open-source deterministic global solver MAiNGO~\cite{MAiNGO}.
%
%
%
%
%
%
%
%
%
%
\newpage
\section{Case study \& results}

In this section, the results of the process optimizations, including the ANN-based surrogate models, are discussed for a specific case scenario (cf. Section~\ref{sec:results_case}). 
In Section~\ref{sec:results_1plantopt} the methodology to integrate multi-scale membrane plant optimization using LbL nanofiltration membranes is presented. 
Then, the performance of the overall process is evaluated when including a more accurate two-dimensional surrogate model by an optimal two-dimensional distribution of the membrane in Section~\ref{sec:results_optimal_solution_strategy}. Here, the single filtration unit is considered by (A) using a one-dimensional distribution, (B)~(1) by using a three-element two-dimensional distribution which is directly implemented in the optimization framework, and (B)~(2) by using a new surrogate model that includes these two-dimensional distributions over $k=1000$ cells in a black box creating a computational resource-efficient two-dimensional modeling of the nanofiltration module. 
In Section~\ref{sec:results_single_salt}, the filtration performance of the process is evaluated when including a second stage. 
Finally, in Section~\ref{sec:results_salt_mixture}, salt mixtures are considered and the process optimized. 
%
%
%
%
%
%
%
%
%
%
\subsection{Case scenario for membrane plant optimization} \label{sec:results_case}

The case scenario and cost correlations are adapted from our previous work~\cite{rall2020simultaneous}. The membrane plant is optimized based on a drinking water purification meeting given quality specifications for a small town with $10,000$ inhabitants~\cite{baur2014mutschmann}. In this study, water softening is considered by retaining \ce{Na2SO4} salt in the membrane plant. The drinking water demand amounts to a peak demand of approximately \SI{224}{\cubic\meter\per\hour}. All processes dimensions are set to meet the permeate volume flow of this peak demand. Demand-side management is not considered within the scope of this work. 
%
%
%
%
%
%
%
%
%
%
\subsection{Bridging the gap of multiple scales by surrogate model-based membrane plant optimization}\label{sec:results_1plantopt}

\begin{figure}[H]
  \centering
  \includegraphics[width=0.85\linewidth]{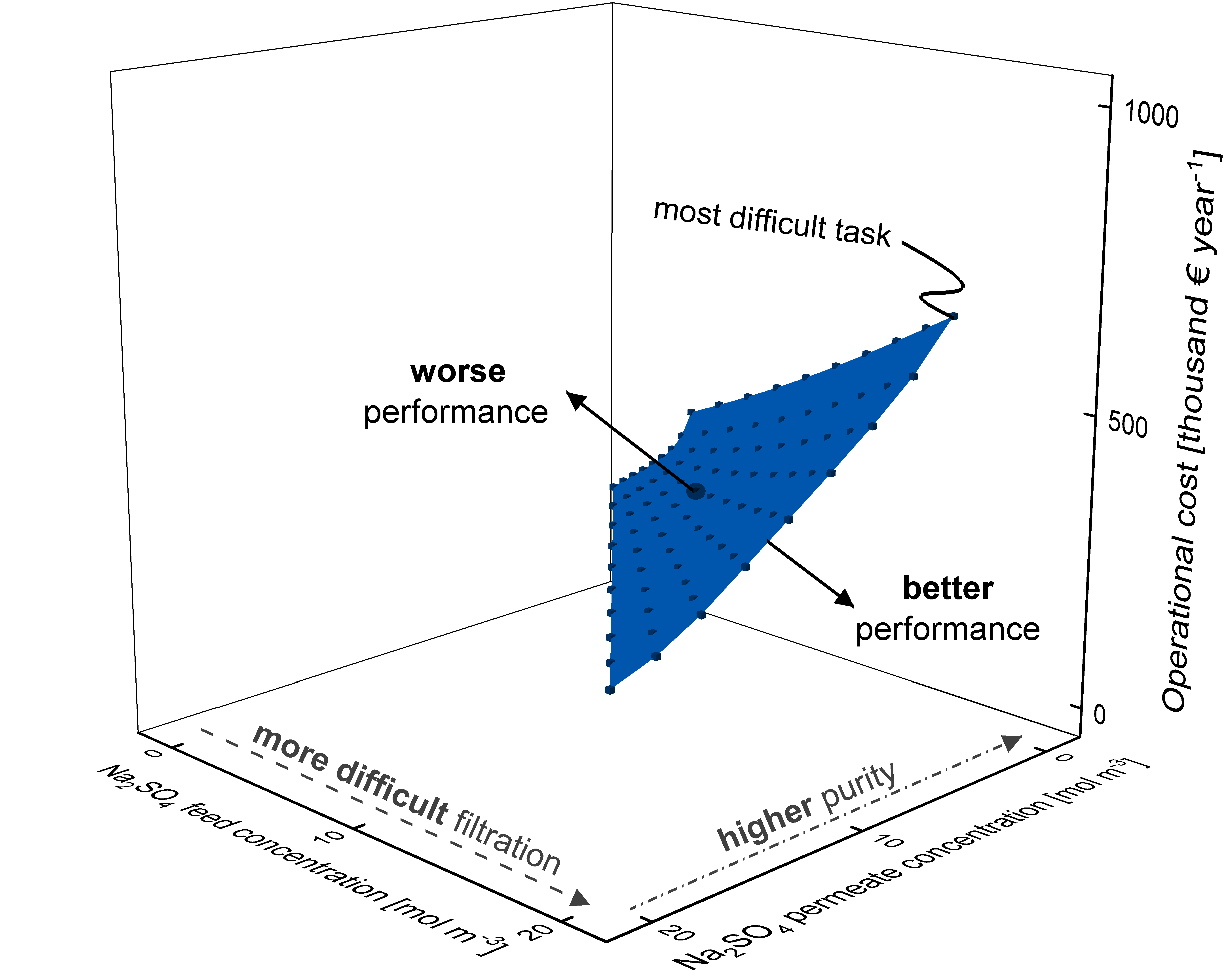}
		\caption{Surrogate model-based membrane plant optimization for a single-stage nanofiltration unit using the ANN$_{Na_{2}SO_{4},~k=1,~1~\text{Unit}}$ surrogate model to describe the ionic retention of the membrane unit for \ce{Na2SO4} feed concentration ranging from \SI{5}{\mol\per\cubic\meter} to \SI{20}{\mol\per\cubic\meter}. The optimization results in a smooth, convex Pareto front. The annual operating costs increase along with higher purity and with the more difficult filtration task.}
		\label{fig:results_1plantopt}
\end{figure}

In this section, the results of the surrogate model-based membrane plant optimization are presented. 
The optimization is performed for a single-stage nanofiltration unit (cf., Figure~\ref{fig:Superstructure}) using the ANN$_{Na_{2}SO_{4},~k=1,~1~\text{Unit}}$ surrogate model (cf., Table~\ref{tab:ArificialNeuralNetworks}) to describe the ionic retention of the membrane unit. The whole process is optimized for ranging feed concentrations of \ce{Na2SO4} and solved for the two objectives -- (i) minimal annual operation costs and (ii) minimal \ce{Na2SO4} permeate concentration. The \ce{Na2SO4} feed concentration range from \SI{5}{\mol\per\cubic\meter} to \SI{20}{\mol\per\cubic\meter}. The results of this optimization are shown in Figure~\ref{fig:results_1plantopt} depicting a smooth, convex Pareto front. 

Ideally, an optimal membrane plant would operate at zero costs and zero permeate concentrations for all feed concentrations. This is called the utopia point. However, in reality, higher feed concentrations (i.e., more difficult filtration) and lower permeate concentrations (i.e., higher purities) always lead to higher operational costs. Hence, we face an inherent trade-off between conflicting objectives. 
This trade-off can be addressed by multi-objective optimization. The solution of a multi-objective optimization problem is a Pareto-front between the objective functions. The points on the Pareto front correspond to processes where neither one of the objectives can be improved anymore without worsening another objective. Thus, for each of the Pareto efficient process points, one cannot archive higher purity at a given feed concentration without spending the additional operational cost. 

Considering the calculated Pareto front in Figure~\ref{fig:results_1plantopt}, indeed the annual operating costs increase along with higher purity (i.e., lower \ce{Na2SO4} permeate concentration) and more difficult filtration task (i.e., higher \ce{Na2SO4} feed concentration concentration). The achievable permeate purity strongly depends on the process' salt feed concentration. Here, the maximum process purity ranges from 0.63~\SI{}{\mol\per\cubic\meter} for a feed concentration of \SI{5}{\mol\per\cubic\meter} to 3.6~\SI{}{\mol\per\cubic\meter} for a feed concentration of \SI{20}{\mol\per\cubic\meter}. The latter is the most difficult separation task of this optimization resulting in an annual operation cost of approximately 700~thousand~\EUR{}~year$^{-1}$. A table of the optimal solution points can be found in the supplementary data to this publication. 

In the presented work, we demonstrate that nano-scale ion transport models can be used in process optimization. Besides, some essential effects for the design of membrane processes result from the observation of the simulated data. Therefore, the qualitative differences between the process configurations are assessed in the following. The tables to this discussion can be found in the supplementary data to this publication. 
The optimal process configuration towards higher purity (and higher annual operation cost) follows a stringent pattern regarding the involved process parameters. The process recirculation (cf., Figure~\ref{fig:Superstructure}~single~stage) is fully exploited to minimize the filtration unit’s feed concentration. Permeate flow recirculation is used to keep the feed concentration low to maintain the membrane's retention high. This process indeed seems to be non-intuitive because the purified permeate stream is recirculated back to the feed stream. However, the feed concentration strongly influences the retention of the nanofiltration membrane and must, therefore, remain low to sustain high ionic retention of the membrane. It becomes apparent that it is economically more viable to recirculate the permeate stream than investing in more membrane area or higher membrane charges associated with low permeability. Moreover, the membrane thickness and surface charge are also increased throughout the optimization to yield a similar, but quantitatively less significant, effect. The increase of the membrane thickness leads to a decreased permeability, which has to be compensated by a larger membrane surface or higher pressure gradient. This can be observed from a large number of membrane modules utilized for high-purity process configurations. Additionally, a large membrane surface facilitates a suitable mass transfer along with a medium transmembrane velocity, which also has a significant impact on retention.

For processes with lower cost and purity, the membrane's surface is drastically reduced to eliminate this cost factor. In particular, no recirculating is performed to minimize the operation costs for the involved feed-sided pressure pump. The preceding observations apply to all following discussed process configurations. For the filtration of salt mixtures, small restrictions have to be made that shall be considered later. 

Furthermore, the shape and position of the calculated Pareto front can be used to determine the performance of the membrane system. In Figure~\ref{fig:results_1plantopt}, a curved Pareto front towards the utopia point indicates overall a better performance. Whereas a curved Pareto away from the utopia point means a worse performance of the process up to the angular end of the feed- and permeate concentration plane so that no filtration takes place. This visual effect may be used when comparing different membrane systems embedded in the same process of plant optimization. Taking a closer look at the next Section~\ref{sec:results_optimal_solution_strategy}, the shape and position of the calculated Pareto front indicates the severe underestimation of the costs when only including a one-dimensional distribution of the membrane.

The optimization problem includes 6 optimization variables, 1 equality constraint, and 9 inequality constraints. The optimization was executed on 4 cores in parallel on a high performance computing cluster. The average CPU time summed over all cores for the optimization of a Pareto point was 2.1 CPU seconds (see Table~\ref{tab:CPUTimeComparison}). Thus, the computation of the Pareto front, which is approximated by 100 points, took roughly 35 CPU minutes in total.
%
%
%
%
%
%
%
%
%
%
\subsection{Optimization including the two-dimensional distribution of the membrane}\label{sec:results_optimal_solution_strategy}

In the next step, the performance of the overall process is evaluated when including a more accurate two-dimensional surrogate model by an optimal two-dimensional distribution of the membrane. A membrane module is considered by (A) using a one-dimensional distribution as presented in the previous Section~\ref{sec:results_1plantopt}), and compared to (B)~(1) using a three-element two-dimensional distribution by three ANN$_{Na_{2}SO_{4},~k=1,~1~\text{Unit}}$ surrogate models (cf., Table~\ref{tab:ArificialNeuralNetworks}) which are directly implemented in the optimization framework, and (B)~(2) by using a new surrogate model ANN$_{Na_{2}SO_{4},~k=1000,~1~\text{Unit}}$ (cf., Table~\ref{tab:ArificialNeuralNetworks2}) that includes these two-dimensional distributions as a black box creating a computational resource-efficient two-dimensional modeling of the nanofiltration module. 

\begin{figure}[H]
  \centering
  \includegraphics[width=1\linewidth]{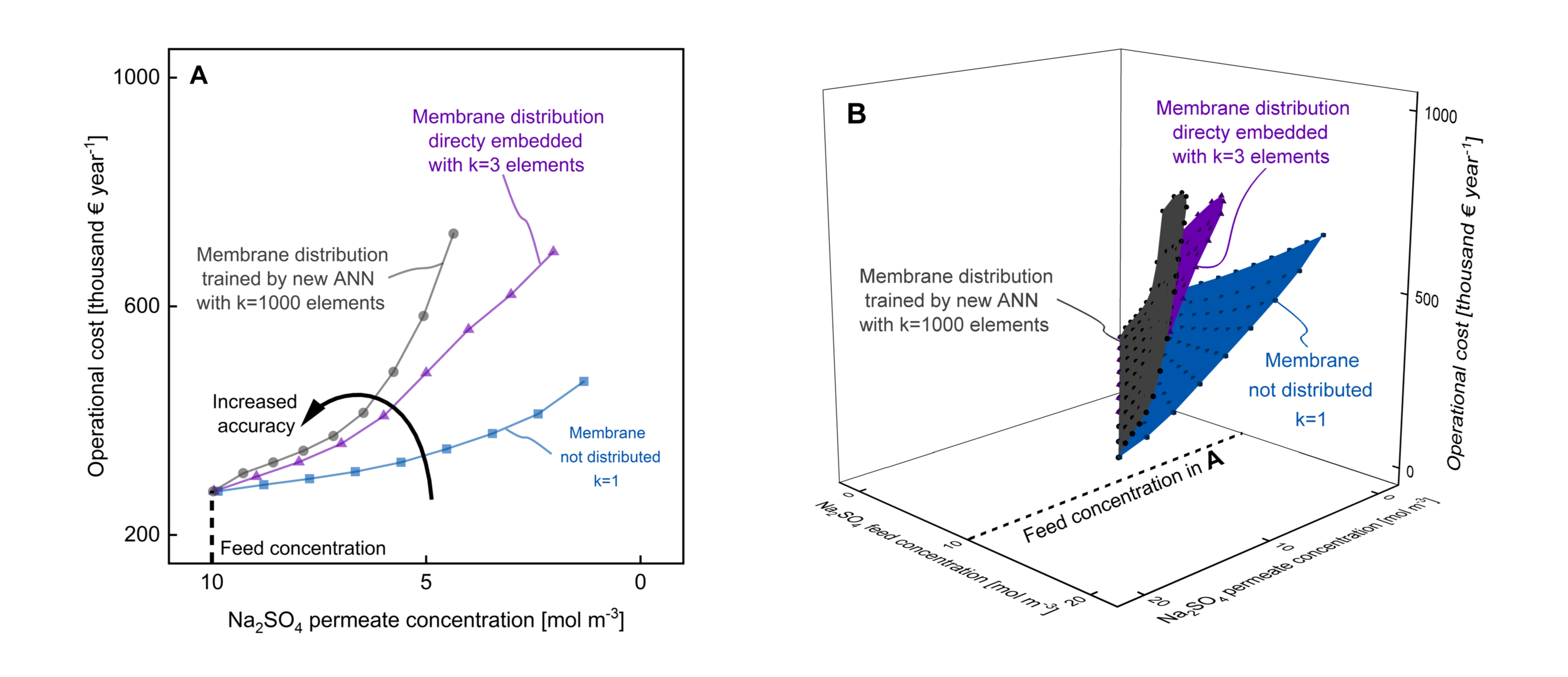}
          \vspace{-1cm}
		\caption{Surrogate model-based membrane plant optimization for a single-stage nanofiltration for \ce{Na2SO4} filtration. Two implementations of the two-dimensional distribution are presented: (B)~(1) by direct implementation of the two-dimensional distribution by three ANN$_{Na_{2}SO_{4},~k=1,~1~\text{Unit}}$ surrogate models (B)~(2) by the recursive back box surrogate model approach using ANN$_{Na_{2}SO_{4},~k=1000,~1~\text{Unit}}$. For comparability, the results of Figure~\ref{fig:results_1plantopt} are displayed using one ANN$_{Na_{2}SO_{4},~k=1,~1~\text{Unit}}$ surrogate model with one-dimensional distribution. Showing A) the results for a feed concentration of \SI{10}{\mol\per\cubic\meter} and B) for ranging feed concentrations between \SI{5}{\mol\per\cubic\meter} and \SI{20}{\mol\per\cubic\meter}.}
		\label{fig:results_ElementsStudy}
\end{figure}

Again, one single-stage membrane unit (c.f., Figure~\ref{fig:Superstructure}) is considered for the optimization of a process plant for ranging feed concentrations of \ce{Na2SO4} and solved for the two objectives - (i) minimal annual operation costs and (ii) minimal \ce{Na2SO4} permeate concentration. The resulting Pareto fronts of the optimizations are shown in Figure~\ref{fig:results_ElementsStudy}. Here, for comparability, the results of Figure~\ref{fig:results_1plantopt} are displayed in Figure~\ref{fig:results_ElementsStudy}~(Membrane~not~distributed, i.e. $k=1$) as well. A table of the optimal solution points can be found in the supplementary data to this publication. 

(B)~(1) Directly embedding the two-dimensional distribution in the optimization framework results in a dramatic increase in run time for only a few two-dimensional distribution elements. 
As shown in Table~\ref{tab:CPUTimeComparison}, the average CPU time to solve one Pareto point using $k=3$ elements is $9.1 \cdot 10^{4}$ CPU seconds corresponding to over 25 CPU hours. To reduce the wall-clock time, we solved this optimization parallel on 48 cores. (Note that the CPU times are summed over all cores.)
Thus, the maximum number of elements for directly embedding the two-dimensional distribution in the optimization framework is $k=3$ elements using a single nanofiltration unit. This low number of elements is not sufficient to accurately describe the changes along the membrane length, as presented in Section~\ref{sec:ANNsMethodsadaption}. The results suggest clearly that the solution of a large number of elements, i.e., $k > 3$, would not be feasible using this method. 

(B)~(2) In contrast, our proposed recursive black-box approach using the ANN$_{Na_{2}SO_{4},~k=1000,~1~\text{Unit}}$ surrogate model requires on average $2.4 \cdot 10^{2}$ CPU seconds to solve a Pareto point (summed over all cores). Thus, the complete Pareto front, including 100 points, is solved within 7 CPU hours. As we use 48 cores in parallel, this corresponds to a wall-clock time of under 10 minutes.
Thus, the method significantly reduced the computational effort compared to (B)~(1) and makes the integration of $k=1000$ elements feasible. Compared to the consideration of $k=1$, we can observe an acceptable increase in CPU time, which is mainly due to the additional input of the ANN describing the flow ratio $\gamma$. 

\begin{table}[h]
    \caption{Computational performance comparison of the proposed optimization methods. For each method, 100 optimization problems are solved, i.e., one for each Pareto point. Not all solved points are displayed in the figure. We provide average, variance, minimal, and maximal CPU times overall optimization problems. Note that CPU times for optimization include preprocessing and branch-and-bound time summed over all parallel cores in seconds.}
	\centering
	\label{tab:CPUTimeComparison}
    \begin{tabular}{lccrr|llll}
        \hline
     & &  &    & &  \multicolumn{4}{c}{CPU time summed over all cores [s]} \\
    \# & Name & Section & k [-] & CPU Cores & Average   & Variance & min & max  \\
        \hline
    A & ${Na_{2}SO_{4},~k=1,~ 1~\text{Unit}}$  & \ref{sec:results_1plantopt}     & 1 & 4 & $2.1 \cdot 10^{1}$ & $2.4 \cdot 10^{2}$ & $7.7 \cdot 10^{-1}$ & $5.6 \cdot 10^{1}$  \\
    B~(1)& ${Na_{2}SO_{4},~k=3,~ 1~\text{Unit}}$   & \ref{sec:results_optimal_solution_strategy} & 3 & 48 & $9.1 \cdot 10^{4}$ & $2.5 \cdot 10^{10}$ & $2.3 \cdot 10^{0}$ & $7.7 \cdot 10^{5}$ \\
    B~(2)& ${Na_{2}SO_{4},~k=1000,~ 1~\text{Unit}}$ & \ref{sec:results_optimal_solution_strategy} & 1000 & 48 & $2.4 \cdot 10^{2}$ & $1.1 \cdot 10^{5}$ & $1.7 \cdot 10^{0}$ & $1.5 \cdot 10^{3}$ \\
        \hline
    \end{tabular}
\end{table}

A severe underestimation of the costs is observed in Figure~\ref{fig:results_ElementsStudy}~A when not including a two-dimensional distribution of the membrane. This severe underestimation of the costs is prevailing for all feed concentration of \ce{Na2SO4} from \SI{5}{\mol\per\cubic\meter} to \SI{20}{\mol\per\cubic\meter} as indicated by the location of the three Pareto surfaces shown in Figure~\ref{fig:results_ElementsStudy}~B. The use of a more precise two-dimensional distribution of the membrane reveals a worse performance than the idealized process only with one-dimensional distribution in Figure~\ref{fig:results_ElementsStudy}. The worsening in process performance can be explained by the filtration-related decreasing retention of salt at the retentate side. For positive retention (which is always the case for the single salt filtration examined here), the concentration on the retentate side of the filtration unit steadily increases along the membrane surface. However, this leads to decreasing retention and, therefore, a worse filtration performance. Additionally, a back-coupling through recirculation of the permeate feed stream back to the membrane's unit feed stream is observed. For the two-dimensional distribution of the membrane with $k=1000$ elements, for instance, the maximum achievable performance for a feed concentration of \SI{5}{\mol\per\cubic\meter} is 1.19~\SI{}{\mol\per\cubic\meter} and therefore presents an increase in impurity as compared to the idealized case using the one-dimensional distribution. This effect is even more pronounced when considering higher feed concentrations. 
%
%
%
%
%
%
%
%
%
%
\subsection{Extension of the optimization process by a second filtration unit}\label{sec:results_single_salt}

Next, the two-dimensional surrogate model approach is utilized to extend the optimization process by a second filtration unit (cf,. Figure~\ref{fig:Superstructure}~two~stage). The ANN$_{Na_{2}SO_{4},~k=1000,~1~\text{Unit}}$ surrogate model is taken from the previous Section~\ref{sec:results_optimal_solution_strategy}. For the two-stage process the ANN$_{Na_{2}SO_{4},~k=1000,~2~\text{Units}}$ surrogate model is chosen (cf., Table~\ref{tab:ArificialNeuralNetworks2}). The whole process is optimized for ranging feed concentrations of \ce{Na2SO4} from \SI{5}{\mol\per\cubic\meter} to \SI{20}{\mol\per\cubic\meter} and solved again for the two objectives - (i) minimal annual operation costs and (ii) minimal \ce{Na2SO4} permeate concentration. The results of the optimization are shown in Figure~\ref{fig:results_MultibleUnits}. A table of the optimal solution points can be found in the supplementary data to this publication. 

\begin{figure}[H]
  \centering
  \includegraphics[width=1\linewidth]{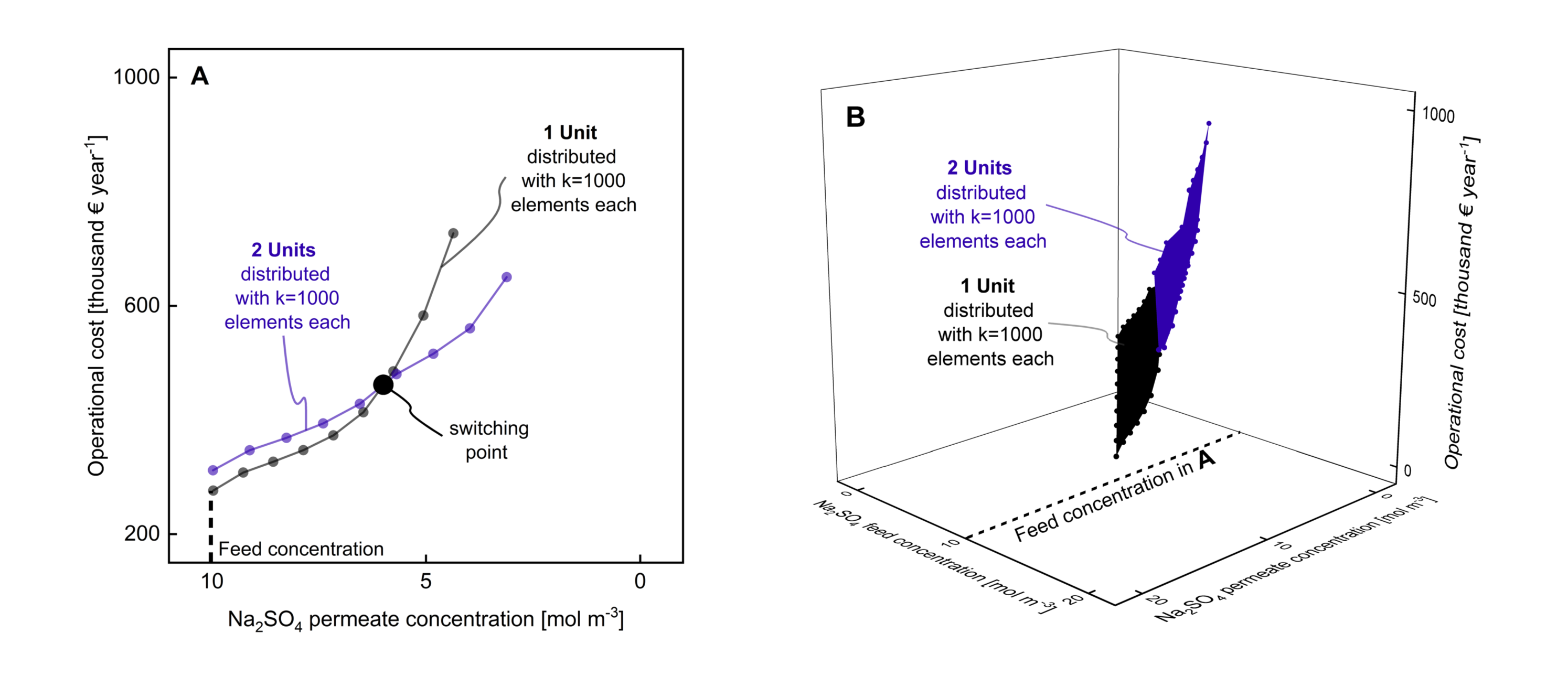}
  \vspace{-1cm}
		\caption{Surrogate model-based membrane plant optimization for a single-stage and two-stage serial connection of nanofiltration units for \ce{Na2SO4} filtration. The two-dimensional surrogate model approach approach is chosen using ANN$_{Na_{2}SO_{4},~k=1000,~1~\text{Unit}}$ for one nanofiltration unit, and the ANN$_{Na_{2}SO_{4},~k=1000,~2~\text{Units}}$ surrogate model for the two-stage serial connection. Showing A) the results for a feed concentration of \SI{10}{\mol\per\cubic\meter} and B) for ranging feed concentrations between \SI{5}{\mol\per\cubic\meter} and \SI{20}{\mol\per\cubic\meter}.}
		\label{fig:results_MultibleUnits}
\end{figure}

This combination of a two-stage series of filtration units results in a strongly enhanced filtration performance. This tendency is particularly intensified as the feed concentration of the second unit is already the purified concentration of the first unit. Due to the strong correlation of the feed concentration and retention, the second filtration unit is particularly useful. Notably, the improvement is much more significant for lower feed concentrations. For higher initial feed concentrations, however, the effect of utilizing two filtration units is not as predominant.

There exists a distinct switching point, as highlighted in Figure~\ref{fig:results_MultibleUnits}~A, where the economic viability of either one or two nanofiltration unit changes. In the range of less difficult filtration operations, a single-stage option is preferable. More difficult filtration operations with higher purity, however, are cheaper under the inclusion of a second unit (or even impossible for a single unit). The costs of a second filtration unit fall short in the overall economic estimation. This distinct switching point is case-specific and dependent on the feed concentration of \ce{Na2SO4}. As shown in Figure~\ref{fig:results_MultibleUnits}~B for all feed concentration of \ce{Na2SO4} from \SI{5}{\mol\per\cubic\meter} to \SI{20}{\mol\per\cubic\meter} a distinct switching point exists highlighted by the change of color (here from black to purple). 

The superstructure optimization includes 12 continuous and 2 integer optimization variables, 6 equality constraints, and 14 inequality constraints. We run the optimization for the two-unit superstructure with $k=1$ elements per unit on 48 cores (result not displayed in Figure~\ref{fig:results_MultibleUnits}). Here, the average CPU time summed over all cores for one Pareto point was $7.7 \cdot 10^{3}$ seconds. This is an increase by a factor of 360 compared to the one unit process with a $k=1$ element.
Due to the increased computational effort, the results with two units and $k=1000$ have been solved using a local optimization approach with 10 multistarts. This method cannot guarantee global optimality. Therefore, only the results with two units and $k=1$ can guarantee global optimality, but are not meaningful as they severely underestimate the overall process costs. The local optimization took on average 25 CPU seconds for a Pareto point.
%
%
%
%
%
%
%
%
%
%
\subsection{Extension of the optimization process by salt mixtures}\label{sec:results_salt_mixture}

\begin{figure}[H]
  \centering
  \includegraphics[width=0.9\linewidth]{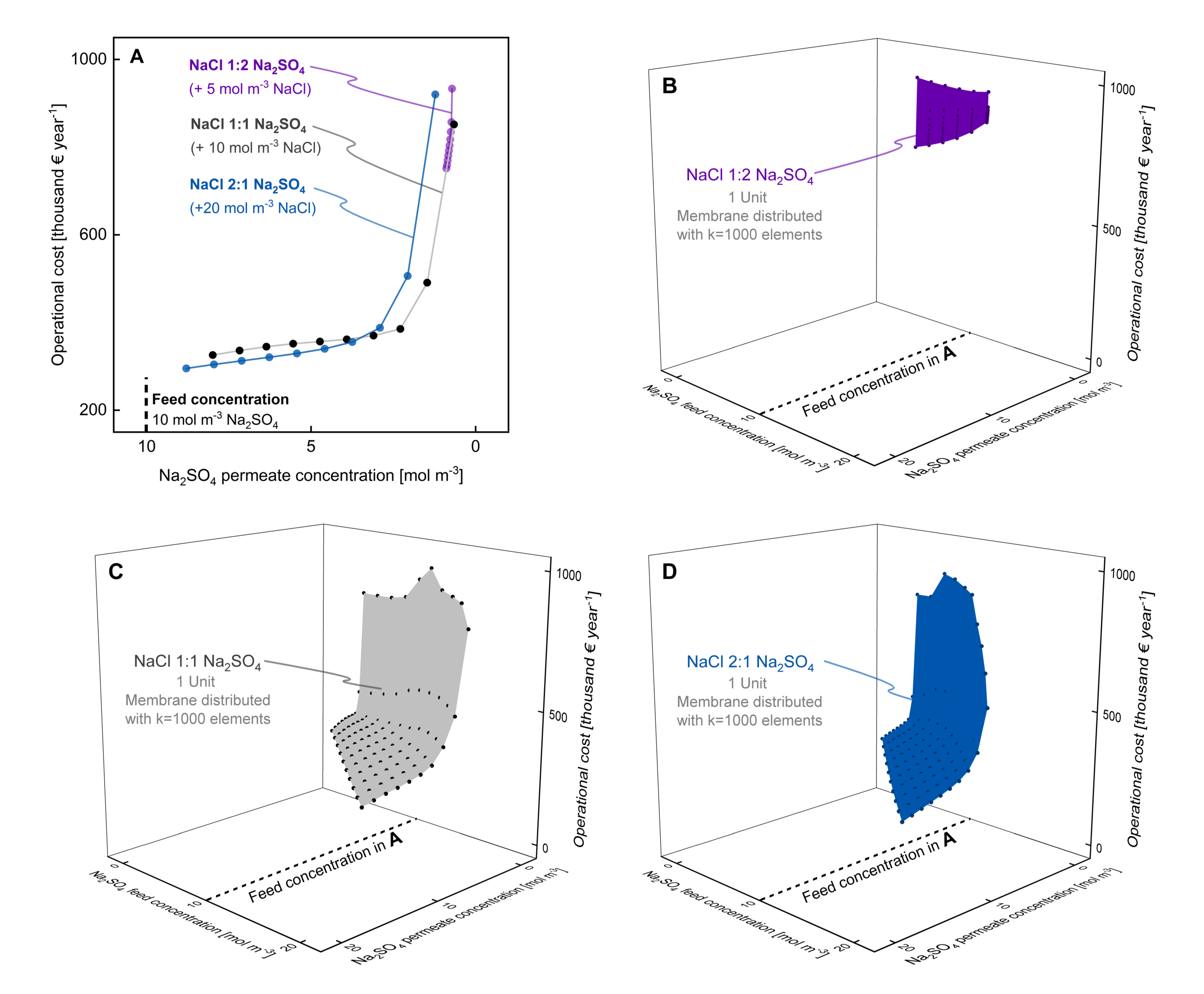}
		\caption{Surrogate model-based membrane plant optimization for a single-stage nanofiltration unit for filtration of \ce{NaCl} and \ce{Na2SO4} salt mixture. The two-dimensional surrogate model approach is chosen using the ANN$_{NaCl,~Na_{2}SO_{4},~k=1000,~1~\text{Unit}}$ surrogate model. Showing A) the results for a \ce{Na2SO4} feed concentration of \SI{10}{\mol\per\cubic\meter} and for ranging feed concentrations between \SI{5}{\mol\per\cubic\meter} and \SI{20}{\mol\per\cubic\meter} for salt mixture ratios (\ce{NaCl}:\ce{Na2SO4}) of B) 1:2, C) 1:1, and D) 2:1.}
		\label{fig:results_salt_mixture}
\end{figure}

Next, the two-dimensional surrogate model approach is extended by salt mixtures. Due to computational effort, only one single-stage is considered in the optimization (cf,. Figure~\ref{fig:Superstructure}). The ANN$_{NaCl,~Na_{2}SO_{4},~k=1000,~1~\text{Unit}}$ surrogate model is utilized in the process optimization framework. The whole process is optimized for ranging feed concentrations of \ce{Na2SO4} from \SI{5}{\mol\per\cubic\meter} to \SI{20}{\mol\per\cubic\meter}. The feed concentration is extended by the addition of \ce{NaCl} by fixed composition ratios. More precise the ratios for \ce{NaCl}:\ce{Na2SO4} are 1:2, 1:1, and 2:1. For example in Figure~\ref{fig:results_salt_mixture}~A, for the ratio \ce{NaCl}:\ce{Na2SO4} = 1:2, \SI{5}{\mol\per\cubic\meter} of \ce{NaCl} is added to the feed stream when considering a feed of \SI{10}{\mol\per\cubic\meter} \ce{Na2SO4}. 
The optimization is again solved for the two objectives - (i) minimal annual operation costs and (ii) minimal \ce{Na2SO4} permeate concentration. The results of the optimization for the given ratios for \ce{NaCl}:\ce{Na2SO4} are shown in Figure~\ref{fig:results_salt_mixture}. A table of the optimal solution points can be found in the supplementary data to this publication. 

To reduce computational efforts, the optimization problem is solved for fewer operating conditions (Pareto pints), leading to a more coarse approximation of the Pareto front. The shape and different process configurations show qualitatively similar behavior concerning the optimization of the single salt processes in Figure~\ref{fig:results_ElementsStudy}. In particular, the operational costs are in the same order of magnitude for both cases.

Taking a closer look at the different Pareto fronts for the individual salt mixtures, they reveal differences in filtration operation. On the one hand, the Pareto fronts for ratios 1:1 and 2:1 (\ce{NaCl}:\ce{Na2SO4}) in Figure~\ref{fig:results_salt_mixture}~C-D present similar behavior to the previous results. On the other hand, the optimization points for a ratio of 1:2 in Figure~\ref{fig:results_salt_mixture}~B are clustered in a much narrower region of feasible points in the Pareto front. This process infeasibility is mainly observed for higher feed concentrations. This behavior can be ascribed to the more complex relations between the membrane retention, and it’s influencing process-specific parameters. These relations are neither monotone (as for the single salt case), nor is the retention itself limited to positive values (the Donnan effect induces negative retention). Most notably filtration operation with high retention is still possible in Figure~\ref{fig:results_salt_mixture}~D even at very high salt concentrations (i.e., \SI{40}{\mol\per\cubic\meter} of \ce{Na2SO4} and \SI{20}{\mol\per\cubic\meter} of \ce{Na2SO4}) by considering the Donnan effect. Hence an accurate description of the underlying physical processes becomes quite cumbersome, making the surrogate-based model for the determination of optimal operation points even more valuable. The increased complexity of the process for salt mixtures also manifests in clustered points on the Pareto front. Such points are Pareto-dominated by preceding configurations, which means a process in that particular setting gets more expensive under looser purity constraints.

The optimization for salt mixtures includes 6 continuous and 1 integer optimization variables, 2 equality constraints, and 13 inequality constraints. We run the optimization for one single-stage unit and $k=1000$ elements on 48 cores. The average CPU time summed over all cores for one Pareto point was $3.3 \cdot 10^{3}$, $1.1 \cdot 10^{5}$, and $1.0 \cdot 10^{4}$ seconds for 1:1, 1:2, and 2:1 salt mixtures (\ce{NaCl}:\ce{Na2SO4}), respectively.
These results demonstrate that the CPU time for optimization is not only dependent on the structure of the optimization problem but also the case study.
%
%
%
%
%
%
%
%
%
%
\newpage
\section{Conclusion}

We present a computational resource-efficient methodology to integrate multi-scale modeling in membrane science. We propose a machine learning approach to integrate accurate physical models of ion transport, valid on the nano-scale, into large-scale superstructure optimization of membrane processes. Thereby, the multi-scale decision-making process allows us to simultaneously design the membrane synthesis properties along with the process design and operation. All models and the solver itself are open-source and freely available, enabling multi-scale modeling in membrane science. 

In particular, we train ANN surrogate models on simulated data from a differential-algebraic one-dimensional extended Nernst-Plank ion transport model, called pEnPEn. These surrogate models are then exploited to extend the framework towards a more accurate two-dimensional distribution of the membrane module to capture the filtration-related decreasing retention of salt. Here, individual ANNs are arranged in series to resolve the membrane in the direction of flow. Next, the two-dimensional surrogate models are embedded in a mechanistic membrane process model leading to a hybrid mechanistic / data-driven model. The results demonstrate that:
\begin{itemize}
    \item A consideration of a one-dimensional distribution of the membrane module leads to a severe underestimation of process costs. Only the extension towards a two-dimensional distribution captures the adverse effects of filtration-related decreasing retention of salt.
    \item Non-intuitive interconnections of the membrane plant layout are optimal for high permeate purities, in which the permeate flow recirculates back to keep the feed concentration low.
    \item A distinct point between a single-stage and a two-stage process exists, where the economic viability of either one or two nanofiltration unit changes. Costs of a second filtration unit fall short in the overall economic estimation for more difficult filtration operations. 
    \item For salt mixtures, it is much more complex to find optimal operating points, as further effects (i.e., Donnan exclusion) have an additional impact on the separation process. Therefore, it is all the more important that surrogate-based modeling bridges the gap between the scales for application of high fidelity ion transport models in superstructure optimization of membrane processes. 
\end{itemize}

This work stimulates to exploit this open-source optimization framework by further research and in industrial applications, such as in drinking water treatment and water disposal. Additional surrogate models may include other membrane-based platforms. Furthermore, the concept of surrogate-based modeling should be exploited to integrate driving force diminishing effects acting on nano-scale, such as the osmotic pressure difference. Future work in deterministic global optimization and model reduction is desirable to push the computational limits of this approach further.
%
%
%
%
%
%
%
%
%
%
\vspace{0.5cm}
\section*{Acknowledgement}
D.R., E.E., and M.W. acknowledge the support through the German Federal Ministry of Education and Research (BMBF) under the project UO-Rohrfabrikation (03XP0100E) and the project EfflueNF (02WIL1486) and the European Union's Horizon~2020 research and innovation program (grant agreement no. 694946). 
A.M.S. and A.M. gratefully acknowledge funding by the excellence initiative of the German federal and state governments and they acknowledge the financial support of the Kopernikus project SynErgie by the Federal Ministry of Education and Research (BMBF) and the project supervision by the project management organization Projekttr\"ager J\"ulich (PtJ).
Funded by the Excellence Initiative of the German federal and state governments. Simulations were performed with computing resources granted by RWTH Aachen University under project rwth0404.

\vspace{0.5 cm}
\section*{References}
\bibliographystyle{elsarticle-num}
\biboptions{sort&compress}
\bibliography{Publication1}{}

\begin{thebibliography}{10}
\expandafter\ifx\csname url\endcsname\relax
  \def\url#1{\texttt{#1}}\fi
\expandafter\ifx\csname urlprefix\endcsname\relax\def\urlprefix{URL }\fi
\expandafter\ifx\csname href\endcsname\relax
  \def\href#1#2{#2} \def\path#1{#1}\fi

\bibitem{larsen2016emerging}
T.~A. Larsen, S.~Hoffmann, C.~L{\"u}thi, B.~Truffer, M.~Maurer, Emerging
  solutions to the water challenges of an urbanizing world, Science 352~(6288)
  (2016) 928--933.

\bibitem{eggimann2017potential}
S.~Eggimann, L.~Mutzner, O.~Wani, M.~Y. Schneider, D.~Spuhler, M.~Moy~de Vitry,
  P.~Beutler, M.~Maurer, The potential of knowing more: A review of data-driven
  urban water management, Environmental Science \& Technology 51~(5) (2017)
  2538--2553.

\bibitem{al2019can}
S.~Al~Aani, T.~Bonny, S.~W. Hasan, N.~Hilal, Can machine language and
  artificial intelligence revolutionize process automation for water treatment
  and desalination?, Desalination 458 (2019) 84--96.

\bibitem{bagheri2019advanced}
M.~Bagheri, A.~Akbari, S.~A. Mirbagheri, Advanced control of membrane fouling
  in filtration systems using artificial intelligence and machine learning
  techniques: A critical review, Process Safety and Environmental Protection
  123 (2019) 229--252.

\bibitem{nunes2019thinking}
S.~P. Nunes, P.~Z. Culfaz-Emecen, G.~Z. Ramon, T.~Visser, G.~H. Koops, W.~Jin,
  M.~Ulbricht, Thinking the future of membranes: Perspectives for advanced and
  new membrane materials and manufacturing processes, Journal of Membrane
  Science 598 (2020) 117761.

\bibitem{ghaffour2013technical}
N.~Ghaffour, T.~M. Missimer, G.~L. Amy, Technical review and evaluation of the
  economics of water desalination: current and future challenges for better
  water supply sustainability, Desalination 309 (2013) 197--207.

\bibitem{abels2013membrane}
C.~Abels, F.~Carstensen, M.~Wessling, Membrane processes in biorefinery
  applications, Journal of Membrane Science 444 (2013) 285--317.

\bibitem{niewersch2014nanofiltration}
C.~Niewersch, A.~B. Bloch, S.~Y{\"u}ce, T.~Melin, M.~Wessling, Nanofiltration
  for the recovery of phosphorus -— development of a mass transport model,
  Desalination 346 (2014) 70--78.

\bibitem{nair2018membrane}
R.~R. Nair, E.~Protasova, S.~Strand, T.~Bilstad, Membrane performance analysis
  for smart water production for enhanced oil recovery in carbonate and
  sandstone reservoirs, Energy \& Fuels 32~(4) (2018) 4988--4995.

\bibitem{werber2016materials}
J.~R. Werber, C.~O. Osuji, M.~Elimelech, Materials for next-generation
  desalination and water purification membranes, Nature Reviews Materials 1~(5)
  (2016) 16018.

\bibitem{luo2018selectivity}
T.~Luo, S.~Abdu, M.~Wessling, Selectivity of ion exchange membranes: A review,
  Journal of Membrane Science 555 (2018) 429--454.

\bibitem{shannon2008science}
M.~A. Shannon, P.~W. Bohn, M.~Elimelech, J.~G. Georgiadis, B.~J. Marinas, A.~M.
  Mayes, Science and technology for water purification in the coming decades,
  Nature 452~(7185) (2008) 301.

\bibitem{remmen2019phosphorus}
K.~Remmen, B.~M{\"u}ller, J.~K{\"o}ser, M.~Wessling, T.~Wintgens, Phosphorus
  recovery in an acidic environment using layer-by-layer modified membranes,
  Journal of Membrane Science 582 (2019) 254--263.

\bibitem{liu2018porous}
W.~Liu, S.~Wijeratne, L.~Yang, M.~Bruening, Porous star-star polyelectrolyte
  multilayers for protein binding, Polymer 138 (2018) 267--274.

\bibitem{harris2000layered}
J.~J. Harris, J.~L. Stair, M.~L. Bruening, Layered polyelectrolyte films as
  selective, ultrathin barriers for anion transport, Chemistry of Materials
  12~(7) (2000) 1941--1946.

\bibitem{malaisamy2005high}
R.~Malaisamy, M.~L. Bruening, High-flux nanofiltration membranes prepared by
  adsorption of multilayer polyelectrolyte membranes on polymeric supports,
  Langmuir 21~(23) (2005) 10587--10592.

\bibitem{cheng2018selective}
W.~Cheng, C.~Liu, T.~Tong, R.~Epsztein, M.~Sun, R.~Verduzco, J.~Ma,
  M.~Elimelech, Selective removal of divalent cations by polyelectrolyte
  multilayer nanofiltration membrane: role of polyelectrolyte charge, ion size,
  and ionic strength, Journal of Membrane Science 559 (2018) 98--106.

\bibitem{ilyas2017preparation}
S.~Ilyas, R.~English, P.~Aimar, J.-F. Lahitte, W.~M. De~Vos, Preparation of
  multifunctional hollow fiber nanofiltration membranes by dynamic assembly of
  weak polyelectrolyte multilayers, Colloids and Surfaces A: Physicochemical
  and Engineering Aspects 533 (2017) 286--295.

\bibitem{menne2016regenerable}
D.~Menne, C.~{\"U}z{\"u}m, A.~Koppelmann, J.~E. Wong, C.~van Foeken, F.~Borre,
  L.~D{\"a}hne, T.~Laakso, A.~Pihlajam{\"a}ki, M.~Wessling, Regenerable
  polymer/ceramic hybrid nanofiltration membrane based on polyelectrolyte
  assembly by layer-by-layer technique, Journal of Membrane Science 520 (2016)
  924--932.

\bibitem{menne2016precise}
D.~Menne, J.~Kamp, J.~E. Wong, M.~Wessling, Precise tuning of salt retention of
  backwashable polyelectrolyte multilayer hollow fiber nanofiltration
  membranes, Journal of Membrane Science 499 (2016) 396--405.

\bibitem{rall2019rational}
D.~Rall, D.~Menne, A.~M. Schweidtmann, J.~Kamp, L.~von Kolzenberg, A.~Mitsos,
  M.~Wessling, Rational design of ion separation membranes, Journal of Membrane
  Science 569 (2019) 209--219.

\bibitem{labban2018relating}
O.~Labban, C.~Liu, T.~H. Chong, J.~H. Lienhard, Relating transport modeling to
  nanofiltration membrane fabrication: navigating the permeability-selectivity
  trade-off in desalination pretreatment, Journal of Membrane Science 554
  (2018) 26--38.

\bibitem{dirir2014theoretical}
Y.~I. Dirir, Y.~Hanafi, A.~Ghoufi, A.~Szymczyk, Theoretical investigation of
  the ionic selectivity of polyelectrolyte multilayer membranes in
  nanofiltration, Langmuir 31~(1) (2014) 451--457.

\bibitem{rall2020simultaneous}
D.~Rall, A.~M. Schweidtmann, B.~M. Aumeier, J.~Kamp, J.~Karwe, K.~Ostendorf,
  A.~Mitsos, M.~Wessling, Simultaneous rational design of ion separation
  membranes and processes, Journal of Membrane Science 600 (2020) 117860.

\bibitem{lonsdale1965transport}
H.~Lonsdale, U.~Merten, R.~Riley, Transport properties of cellulose acetate
  osmotic membranes, Journal of Applied Polymer Science 9~(4) (1965)
  1341--1362.

\bibitem{schlogl1966membrane}
R.~Schl{\"o}gl, Membrane permeation in systems far from equilibrium, Berichte
  der Bunsengesellschaft f{\"u}r Physikalische Chemie 70~(4) (1966) 400--414.

\bibitem{yaroshchuk2013solution}
A.~Yaroshchuk, M.~L. Bruening, E.~E.~L. Bernal,
  Solution-diffusion--electro-migration model and its uses for analysis of
  nanofiltration, pressure-retarded osmosis and forward osmosis in multi-ionic
  solutions, Journal of Membrane Science 447 (2013) 463--476.

\bibitem{femmer2016mechanistic}
R.~Femmer, M.~Mart{\'\i}-Calatayud, M.~Wessling, Mechanistic modeling of the
  dielectric impedance of layered membrane architectures, Journal of Membrane
  Science 520 (2016) 29--36.

\bibitem{bowen2002modelling}
W.~R. Bowen, J.~S. Welfoot, Modelling the performance of membrane
  nanofiltration —- critical assessment and model development, Chemical
  Engineering Science 57~(7) (2002) 1121--1137.

\bibitem{jin2011surrogate}
Y.~Jin, Surrogate-assisted evolutionary computation: Recent advances and future
  challenges, Swarm and Evolutionary Computation 1~(2) (2011) 61--70.

\bibitem{lapkin2010chemical}
A.~A. Lapkin, Chemical engineering science and green chemistry--the challenge
  of sustainability, Handbook of Green Chemistry: Online (2010) 1--16.

\bibitem{chachuat2006global}
B.~Chachuat, A.~B. Singer, P.~I. Barton, Global methods for dynamic
  optimization and mixed-integer dynamic optimization, Industrial \&
  Engineering Chemistry Research 45~(25) (2006) 8373--8392.

\bibitem{singer2006global}
A.~B. Singer, P.~I. Barton, Global optimization with nonlinear ordinary
  differential equations, Journal of Global Optimization 34~(2) (2006)
  159--190.

\bibitem{wesselhoeft2018algorithms}
C.~Wesselhoeft, D.~A. Ham, R.~Misener, Algorithms for mixed-integer
  optimization constrained by partial differential equations, in: Computer
  Aided Chemical Engineering, Vol.~44, Elsevier, 2018, pp. 799--804.

\bibitem{sager2015efficient}
S.~Sager, M.~Claeys, F.~Messine, Efficient upper and lower bounds for global
  mixed-integer optimal control, Journal of Global Optimization 61~(4) (2015)
  721--743.

\bibitem{ohs2016optimization}
B.~Ohs, J.~Lohaus, M.~Wessling, Optimization of membrane based nitrogen removal
  from natural gas, Journal of Membrane Science 498 (2016) 291--301.

\bibitem{zarca2019optimization}
R.~Zarca, A.~Ortiz, D.~Gorri, L.~T. Biegler, I.~Ortiz, Optimization of
  multistage olefin/paraffin membrane separation processes through rigorous
  modeling, AIChE Journal 65~(6) (2019) e16588.

\bibitem{lee2018automated}
S.~Lee, M.~Binns, J.-K. Kim, Automated process design and optimization of
  membrane-based {CO2} capture for a coal-based power plant, Journal of
  Membrane Science 563 (2018) 820--834.

\bibitem{bocking2019can}
A.~B{\"o}cking, V.~Koleva, J.~Wind, Y.~Thiermeyer, S.~Blumenschein, R.~Goebel,
  M.~Skiborowski, M.~Wessling, Can the variance in membrane performance
  influence the design of organic solvent nanofiltration processes?, Journal of
  Membrane Science 575 (2019) 217--228.

\bibitem{mores2018membrane}
P.~Mores, A.~Arias, N.~Scenna, J.~Caballero, S.~Mussati, M.~Mussati,
  Membrane-based processes: optimization of hydrogen separation by minimization
  of power, membrane area, and cost, Processes 6~(11) (2018) 221.

\bibitem{alsayegh2017systematic}
S.~Alsayegh, J.~Johnson, B.~Ohs, J.~Lohaus, M.~Wessling, {Systematic
  optimization of H2 recovery from water splitting process using membranes and
  N2 diluent}, International Journal of Hydrogen Energy 42~(9) (2017)
  6000--6011.

\bibitem{ghobeity2014optimal}
A.~Ghobeity, A.~Mitsos, Optimal design and operation of desalination systems:
  new challenges and recent advances, Current Opinion in Chemical Engineering 6
  (2014) 61--68.

\bibitem{white2019multiscale}
D.~A. White, W.~J. Arrighi, J.~Kudo, S.~E. Watts, Multiscale topology
  optimization using neural network surrogate models, Computer Methods in
  Applied Mechanics and Engineering 346 (2019) 1118--1135.

\bibitem{von2014hybrid}
M.~von Stosch, R.~Oliveira, J.~Peres, S.~F. de~Azevedo, Hybrid semi-parametric
  modeling in process systems engineering: past, present and future, Computers
  \& Chemical Engineering 60 (2014) 86--101.

\bibitem{zhou2019big}
T.~Zhou, Z.~Song, K.~Sundmacher, Big data creates new opportunities for
  materials research: a review on methods and applications of machine learning
  for materials design, Engineering 5~(6) (2019) 1017 -- 1026.

\bibitem{tsay2019110th}
C.~Tsay, M.~Baldea, 110th anniversary: using data to bridge the time and length
  scales of process systems, Industrial \& Engineering Chemistry Research
  58~(36) (2019) 16696--16708.

\bibitem{prakash2018chances}
A.~Prakash, S.~Sandfeld, Chances and challenges in fusing data science with
  materials science: the working group {{“3D data science” is headed by
  Prof. Dr. Stefan Sandfeld.}}, Practical Metallography 55~(8) (2018) 493--514.

\bibitem{sanchez2018inverse}
B.~Sanchez-Lengeling, A.~Aspuru-Guzik, Inverse molecular design using machine
  learning: generative models for matter engineering, Science 361~(6400) (2018)
  360--365.

\bibitem{henao2011surrogate}
C.~A. Henao, C.~T. Maravelias, Surrogate-based superstructure optimization
  framework, AIChE Journal 57~(5) (2011) 1216--1232.

\bibitem{unger2009neural}
J.~F. Unger, C.~K{\"o}nke, Neural networks as material models within a
  multiscale approach, Computers \& Structures 87~(19-20) (2009) 1177--1186.

\bibitem{lee2018machine}
J.~H. Lee, J.~Shin, M.~J. Realff, Machine learning: overview of the recent
  progresses and implications for the process systems engineering field,
  Computers \& Chemical Engineering 114 (2018) 111--121.

\bibitem{venkatasubramanian2019promise}
V.~Venkatasubramanian, The promise of artificial intelligence in chemical
  engineering: is it here, finally, AIChE J 65~(2) (2019) 466--78.

\bibitem{mistry2018mixed}
M.~Mistry, D.~Letsios, G.~Krennrich, R.~M. Lee, R.~Misener, Mixed-integer
  convex nonlinear optimization with gradient-boosted trees embedded, arXiv
  preprint arXiv:1803.00952 (2018).

\bibitem{boukouvala2016data}
F.~Boukouvala, J.~Li, X.~Xiao, C.~A. Floudas, Data-driven modeling and global
  optimization of industrial-scale petrochemical planning operations, in: 2016
  American Control Conference (ACC), IEEE, 2016, pp. 3340--3345.

\bibitem{cozad2014learning}
A.~Cozad, N.~V. Sahinidis, D.~C. Miller, Learning surrogate models for
  simulation-based optimization, AIChE Journal 60~(6) (2014) 2211--2227.

\bibitem{fahmi2012process}
I.~Fahmi, S.~Cremaschi, Process synthesis of biodiesel production plant using
  artificial neural networks as the surrogate models, Computers \& Chemical
  Engineering 46 (2012) 105--123.

\bibitem{del2019review}
E.~A. Del Rio-Chanona, X.~Cong, E.~Bradford, D.~Zhang, K.~Jing, Review of
  advanced physical and data-driven models for dynamic bioprocess simulation:
  case study of algae--bacteria consortium wastewater treatment, Biotechnology
  and Bioengineering 116~(2) (2019) 342--353.

\bibitem{schafer2019reduced}
P.~Sch{\"a}fer, A.~Caspari, K.~Kleinhans, A.~Mhamdi, A.~Mitsos, Reduced dynamic
  modeling approach for rectification columns based on compartmentalization and
  artificial neural networks, AIChE Journal 65~(5) (2019).

\bibitem{schweidtmann2019deterministic}
A.~M. Schweidtmann, A.~Mitsos, Deterministic global optimization with
  artificial neural networks embedded, Journal of Optimization Theory and
  Applications 180~(3) (2019) 925--948.

\bibitem{schweidtmann2019singlespecies}
A.~M. Schweidtmann, W.~R. Huster, J.~T. L{\"u}thje, A.~Mitsos, Deterministic
  global process optimization: accurate (single-species) properties via
  artificial neural networks, Computers \& Chemical Engineering 121 (2019)
  67--74.

\bibitem{huster2019WorkingFluidSelection}
W.~R. Huster, A.~M. Schweidtmann, A.~Mitsos, Working fluid selection for
  organic rankine cycles via deterministic global optimization of design and
  operation, {Optimization \& Engineering} (2019) 1--20.

\bibitem{schweidtmann2019flash}
A.~M. Schweidtmann, D.~Bongartz, W.~R. Huster, A.~Mitsos, Deterministic global
  process optimization: flash calculations via artificial neural networks, in:
  Computer Aided Chemical Engineering, Vol.~46, Elsevier, 2019, pp. 937--942.

\bibitem{huster2019impact}
W.~R. Huster, A.~M. Schweidtmann, A.~Mitsos, Impact of accurate working fluid
  properties on the globally optimal design of an organic rankine cycle, in:
  Computer Aided Chemical Engineering, Vol.~47, Elsevier, 2019, pp. 427--432.

\bibitem{schafer2019wavelet}
P.~Sch{\"a}fer, A.~M. Schweidtmann, P.~H. Lenz, H.~M. Markgraf, A.~Mitsos,
  Wavelet-based grid-adaptation for nonlinear scheduling subject to
  time-variable electricity prices, Computers \& Chemical Engineering (2019)
  106598.

\bibitem{madaeni2010modeling}
S.~Madaeni, N.~T. Hasankiadeh, A.~Kurdian, A.~Rahimpour, Modeling and
  optimization of membrane fabrication using artificial neural network and
  genetic algorithm, Separation and Purification Technology 76~(1) (2010)
  33--43.

\bibitem{al2007rejection}
H.~Al-Zoubi, N.~Hilal, N.~Darwish, A.~W. Mohammad, {Rejection and modelling of
  sulphate and potassium salts by nanofiltration membranes: neural network and
  Spiegler--Kedem model}, Desalination 206~(1-3) (2007) 42--60.

\bibitem{wessling1994modelling}
M.~Wessling, M.~Mulder, A.~Bos, M.~Van Der~Linden, M.~Bos, W.~Van Der~Linden,
  Modelling the permeability of polymers: a neural network approach, Journal of
  Membrane Science 86~(1-2) (1994) 193--198.

\bibitem{roehl2018modeling}
E.~A. Roehl~Jr, D.~A. Ladner, R.~C. Daamen, J.~B. Cook, J.~Safarik, D.~W.
  Phipps~Jr, P.~Xie, {Modeling fouling in a large RO system with artificial
  neural networks}, Journal of Membrane Science 552 (2018) 95--106.

\bibitem{salehi2016modeling}
F.~Salehi, S.~M. Razavi, Modeling of waste brine nanofiltration process using
  artificial neural network and adaptive neuro-fuzzy inference system,
  Desalination and Water Treatment 57~(31) (2016) 14369--14378.

\bibitem{soleimani2013experimental}
R.~Soleimani, N.~A. Shoushtari, B.~Mirza, A.~Salahi, Experimental
  investigation, modeling and optimization of membrane separation using
  artificial neural network and multi-objective optimization using genetic
  algorithm, Chemical Engineering Research and Design 91~(5) (2013) 883--903.

\bibitem{femmer2015ion}
R.~Femmer, A.~Mani, M.~Wessling, Ion transport through
  electrolyte/polyelectrolyte multi-layers, Scientific Reports {5} ({2015})
  11583.

\bibitem{evdochenko2019direct}
E.~Evdochenko, J.~Kamp, R.~Femmer, Y.~Xu, V.~Nikonenko, M.~Wessling, Unraveling
  the effect of charge distribution in a polyelectrolyte multilayer
  nanofiltration membrane on its ion transport properties, {Journal of Membrane
  Science} (2020) 118045.

\bibitem{MAiNGO}
D.~Bongartz, J.~Najman, S.~Sass, A.~Mitsos, {MAiNGO} -- {McCormick}-based
  algorithm for mixed-integer nonlinear global optimization, Tech. rep.,
  Process Systems Engineering (AVT.SVT), RWTH Aachen University (2019).

\bibitem{MeLOn_Git}
A.~M. Schweidtmann, L.~Netze, A.~Mitsos, Melon: Machine learning models for
  optimization, \url{https://git.rwth-aachen.de/avt.svt/public/MeLOn/} (2020).

\bibitem{COMSOL}
{COMSOL, Inc}, {COMSOL Multiphysics \textsuperscript{\textregistered} (Version
  5.3)}, \url{https://www.comsol.com}, {accessed 19 June 2019}.

\bibitem{MATLAB2019}
MATLAB, version 9.3 (R2017b), The MathWorks Inc., Natick, Massachusetts, 2017.

\bibitem{lecun2015deep}
Y.~LeCun, Y.~Bengio, G.~Hinton, Deep learning, Nature 521~(7553) (2015)
  436--444.

\bibitem{dayhoff2001artificial}
J.~E. Dayhoff, J.~M. DeLeo, Artificial neural networks: opening the black box,
  Cancer: Interdisciplinary International Journal of the American Cancer
  Society 91~(S8) (2001) 1615--1635.

\bibitem{marquardt1963algorithm}
D.~W. Marquardt, An algorithm for least-squares estimation of nonlinear
  parameters, Journal of the society for industrial and applied mathematics
  11~(2) (1963) 431--441.

\bibitem{yaroshchuk2008negative}
A.~E. Yaroshchuk, Negative rejection of ions in pressure-driven membrane
  processes, Advances in Colloid and Interface Science 139~(1-2) (2008)
  150--173.

\bibitem{Doncevic2020GlobalANNControl}
D.~T. Doncevic, A.~M. Schweidtmann, Y.~Vaupel, P.~Sch{\"a}fer, A.~Caspari,
  A.~Mitsos, Deterministic global nonlinear model predictive control with
  neural networks embedded, Submitted (2020).

\bibitem{pentair}
{Pentair Xflow Products}, \url{https://xflow.pentair.com/en/products/},
  accessed: 1st of April 2019.

\bibitem{dupontUF}
{DU PONT Ultrafiltration Products},
  \url{https://www.dupont.com/water/ultrafiltration.html}, accessed: 1st of
  April 2019.

\bibitem{verberne1993membraanfiltratie}
A.~Verberne, J.~Wouters, Membraanfiltratie voor de drinkwaterbereiding:
  economische optimalisatie van ontwerpparameters, H2O 26~(14) (1993) 383--387.

\bibitem{sethi2000cost}
S.~Sethi, M.~R. Wiesner, Cost modeling and estimation of crossflow membrane
  filtration processes, Environmental Engineering Science 17~(2) (2000) 61--79.

\bibitem{ang2017effect}
W.~Ang, D.~Nordin, A.~Mohammad, A.~Benamor, N.~Hilal, Effect of membrane
  performance including fouling on cost optimization in brackish water
  desalination process, Chemical Engineering Research and Design 117 (2017)
  401--413.

\bibitem{baur2014mutschmann}
A.~Baur, P.~Fritsch, W.~Hoch, G.~Merkl, J.~Rautenberg, M.~Wei{\ss}, B.~Wricke,
  {Mutschmann/Stimmelmayr Taschenbuch der Wasserversorgung} (2014).

\end{thebibliography}

\end{document}